\documentclass[10pt,journal,compsoc]{IEEEtran}

\ifCLASSOPTIONcompsoc

  \usepackage[nocompress]{cite}
\else

  \usepackage{cite}
\fi

\ifCLASSINFOpdf

\else
\fi

\usepackage{balance}  
\usepackage{graphics} 
\usepackage{txfonts}
\usepackage{times}    
\usepackage{color}
\usepackage{textcomp}
\usepackage{booktabs}
\usepackage{ccicons}
\usepackage{todonotes}
\usepackage{subcaption}
\usepackage{graphicx}

\usepackage{caption}

\usepackage{amsmath}

\begin{document}

\title{Tales of Two Cities: Using Social Media to Understand Idiosyncratic Lifestyles in Distinctive Metropolitan Areas}

\author{Tianran~Hu,
        Eric~Bigelow,
        Jiebo~Luo,~\IEEEmembership{Fellow,~IEEE,}
        and~Henry~Kautz,~\IEEEmembership{Member,~IEEE}
\IEEEcompsocitemizethanks{\IEEEcompsocthanksitem Authors are with Computer Science Department, University of Rochester

E-mail: \{thu, jluo, kautz\}@cs.rochester.edu, ebigelow@u.rochester.edu}}

\maketitle

\begin{abstract}
Lifestyles are a valuable model for understanding individuals' physical and mental lives, comparing social groups, and making recommendations for improving people's lives. In this paper, we examine and compare lifestyle behaviors of people living in cities of different sizes, utilizing freely available social media data as a large-scale, low-cost alternative to traditional survey methods. We use the Greater New York City area as a representative for large cities, and the Greater Rochester area as a representative for smaller cities in the United States. We employed matrix factor analysis as an unsupervised method to extract salient mobility and work-rest patterns for a large population of users within each metropolitan area. We discovered interesting human behavior patterns at both a larger scale and a finer granularity than is present in previous literature, some of which allow us to quantitatively compare the behaviors of individuals of living in big cities to those living in small cities. We believe that our social media-based approach to lifestyle analysis represents a powerful tool for social computing in the big data age.
\end{abstract}






\section{Introduction}

\IEEEPARstart{W}{e} take \textit{lifestyle} to be \textit{the way in which a person or group lives including the interests, opinions, behaviors, and behavioral orientations.} Understanding lifestyle is key to gaining insight of the physical and mental aspects of individuals, social groups and cultures. Health, for example, is highly related to one's lifestyle~\cite{cox1987health,lee1999learning}.
Cultural boundaries can be discovered from people's ways of living such as pace of life, eating and drinking habits and so on~\cite{garcia2013cultural,silva2014you}. 
Researchers have also discovered correlations between health and individuals' daily movements as estimated from cellphone GPS tags on social media~\cite{sadilek2013modeling}.
 
In this work, we study the differences of lifestyles in cities of different sizes. A popular stereotype is that life in big cities is fast-paced, high-pressure, and consistently exciting, while life in small cities is calmer and less various due to a lower population density and more limited selection of recreational venues.

We select the Greater New York City area (NYC) as being representative, for our purposes, of big cities in the US. For smaller cities, we select the Great Rochester area (ROC) as representative for two main reasons: First, the size of Rochester (0.2 million) is close to the median size (0.16 million) of cities in the US, approximately 40 times smaller than NYC. Second, these two areas are located close to each other (both in the north-eastern US). Geographic closeness generally leads to similarity of climate and culture, which helps eliminate confounding factors that may lead to differences in lifestyle behaviors unrelated to city size. 

In contrast to traditional research investigating lifestyle patterns, where data collection methods include questionnaires and telephone interviewing~\cite{budesa2008gender,randler2008morningness,singapore}, we leverage data from social media to make inferences about people's lifestyles. The wide adoption of social media brings researchers a new opportunity of studying natural, unconstrained human behavior at very large scales. Foursquare is one of the most popular Location Based Social Networks (LBSNs), holding 5 billion check-in records for 55 million users worldwide\footnote{\url{https://foursquare.com/about}}. This offers us a rich data source for conducting mobility, behavior and lifestyle studies.

We consider temporal and spatial lifestyle in this work. The temporal dimension of a person's lifestyle is assumed to correlate with his/her work-rest ratio in daily actives. In the primary literature on circadian topology (CT), people are classified into one of three categories: morning-types, evening-types, and neither-types ~\cite{horne1975self}. In the CT literature, individuals are modeled by just one of these \textit{types}. In our present work, work-rest behavioral patterns are instead considered to be a weighted combination of all three temporal \textit{lifestyles}: ``Night Owl'', ``Early Bird'' and ``Intermediate''

To avoid assigning a person to a lifestyle in an arbitrary or qualitative fashion, we employ non-negative matrix factorization (NMF) to discover three latent patterns of temporal activity. The extracted patterns offer precise definitions of activity levels associated with specific lifestyles and align with our assumptions about human work-rest habits. A spatial dimension is used to describe lifestyles according to locational behavior. For example, one primitive lifestyle pattern is defined by frequent visits to POIs (points of interest) such as bars and music venues, while another is defined by visits to parks, art galleries and museums. We then apply a clustering method to group these primitive latent patterns into more complex lifestyles that are representative of a group of individuals (e.g. students or stay-at-home parents). We significant variance between the distribution of lifestyles in NYC and ROC. 

Additionally, we use third-order tensor decomposition to find composite patterns across both spatial and temporal dimensions. We extract clearly identifiable patterns of behavior, for example high school students posting during school hours, and for college students frequently visiting or living on campus. This method offers promise as an efficient way of extracting complicated patterns across multiple high-dimensional spaces.

The main contributions of this work are:

1. Use of open-source geo-tagged social media data for analyzing lifestyle patterns as a low-cost, large-scale alternative to traditional survey methods.

2. Application of matrix factor analysis to extract persistent and salient human mobility and work-rest patterns over a large population of users.

3. Application of CP tensor decomposition to discover composite spatial-temporal lifestyle patterns which are useful for understanding fluctuations in people's activity across different time ranges and locations. 

4. Confirming intuitive knowledge and previous research in human activity patterns with quantitative, unsupervised data analysis.

5. Shedding light on the differences and similarities between life in big cities and life in smaller cities, quantitatively confirming many of the common perceptions about life in big and small cities. For example, life in big cities is more work-focused, while it is more home-focused in smaller cities; life in large cities is also more fast-paced and diverse. Furthermore, we have discovered fine-grained lifestyle descriptions that previous small-scale survey-based studies have failed to illuminate. For example, we extracted three types of temporal lifestyles, and report the activity level of each lifestyle along time quantitatively.

\section{Related Work}

\subsection{Sociology and Cronobiology}

Lifestyle is well studied in sociology. The work of~\cite{singapore} suggests that lifestyles such as residential location, mode options, destination choices, and trip timing are constrained by household considerations. Gender difference in lifestyles also attracts interest of many researchers. Budesa et al. study the influence of gender on perceived healthy and unhealthy lifestyles, finding that gender is not an important determinant of individual perceptions about health~\cite{budesa2008gender}. Merritt et al. in~\cite{merritt2003gender} suggest that men and women have no significant difference in motor ability in daily activities. Finally, a study~\cite{von2005gender} on university students finds that female students are healthier due to less alcohol consumption and more healthy habits. 

Much work has been done on human work-rest habits in chronobiology and Circadian Topology (CT). The traditional method of studying how work-rest patterns relate to aspects of physical and mental well-being has been learned through the morningness-eveningness questionnaire (MEQ) of~\cite{horne1976self} and variations of it~\cite{smith1989evaluation}. In the work of Horne et al., Morning-type subjects (MTs) are found to wake at a mean of 7:24am, Neither-type subjects (NTs) at 8:07am, and Evening-type subjects (ETs) at 9:18am; mean bed times for the three types are 11:26pm, 11:30pm, and 1:05am, respectively. These specific times vary in different studies, leading to differing assertions about how much of the population is a member of each CT type~\cite{Taillard1999needsleep, adan2012review}.

Randler finds a significant positive correlation between ``morningness'' tendencies of people and satisfaction in life~\cite{randler2008morningness}, and Monk~\cite{monk2004morningness} find that MT individuals appear to have more regular lifestyle than ETs. A positive correlation between eveningness and depression level is reported by Hasler et al.~\cite{hasler2010morningness}. A thorough review of contemporary CT literature is available in \cite{adan2012review}.

\subsection{Social Media Analytics}

In recent years researchers have successfully utilized social media in research ventures related to lifestyle analysis. Noulas et al. of~\cite{noulas2011empirical} use Foursquare data to discover the behavioral habits of residents in London. The work presented in this paper is strongly inspired by this research: we contribute stacked plots similar to those of Noulas et al., representing the relative visit frequencies of the most frequent POIs, comparing between NYC (\ref{fig:nyc_stacked}) and ROC (\ref{fig:roc_stacked}) and between weekends and weekdays within these cities. Based on the contents of tweets, Sadiek et al. build a language model to detect the health condition of individuals~\cite{sadilek2013nemesis}. By relating a user's health level with other attributes such as environmental features of places where the user spends tags as estimated from his tweet geotags, they estimate the influence of lifestyle on health conditions~\cite{sadilek2013modeling}. Eating and drinking habit is also a key point to understanding human life. In~\cite{abbar2014you} Abbar et al. find out the names of food in people's tweets and use them to estimate the caloric values people possibly take. 

Cranshaw et al.~\cite{cranshaw2010bridging} construct a metric called Location Entropy to measure the diversity of a POI. Sang et al. discuss people's movement session patterns~\cite{sang2015activity} based on China's LBSN data. The eating and drinking habits of different countries and regions are investigated in ~\cite{silva2014you} based on Foursquare data. They find that geographic closeness usually leads to closeness in eating and drinking habits. Wu et al. reported an approach on modeling temporal dynamic in\cite{wu2016unfolding}. Their work showed that besides user-item factors temporal factors are as important in social media popularity prediction. Other aspects of lifestyle such as pace of life and power distance, are discussed in~\cite{garcia2013cultural}. They estimate each index related to life via tweets collected all over the world. In~\cite{golder2011diurnal}, Golder et al. found that the negative affect (NA) tweets sent in winter is higher than those sent in summer.  Similar results are reported in~\cite{park2013mood}, in which the weather influence on human sentiment is studied using tweets. Tensor decomposition was applied in~\cite{zheng2014diagnosing}. In this work, Zheng Yu et. al decomposed third-order tensors to extract noise-location compound patterns in an urban area. We employ similar method in our work to find temporal-spatial compound patterns of lifestyles.

\section{Data Set and Preprocessing}

The large number of self-reported location records and wide geographic coverage make Foursquare a valuable data source for analyzing behavioral tendencies across groups of individuals. However, directly collection of users' check-ins is a nontrivial task due to the strict limits on Foursquare data download rates. As an alternative, many researchers collect Foursquare data through other social media sources that connect with Foursquare such as Twitter~\cite{noulas2011empirical}. If a user links his or her Foursquare account with a Twitter account, when s/he performs a check-in on Foursquare, a geo-tagged tweet will be posted automatically. This tweet contains a link to the webpage of the venue where the user checks in via Foursquare. In the present work, we use this method to collect users' check-in data. To avoid tweets from possible tourists, we filter out users whose tweets appeared exclusively within a period of less than 7 days.

\subsection{Lifestyle Study through Social Media}

We collected 233,046 Foursquare check-ins from 49,744 POIs from geo-tagged tweets in NYC, and 99,466 check-ins from 13,483 POIs from ROC. Foursquare also provides around 600 POI categories, such as Arts \& Entertainment, Home, etc. A venue can be assigned to several categories, where one category can be a subset of another. For example, Foursquare may assign both American Restaurant and Restaurant to a single venue. 

Due to the sparsity of direct Foursquare check-ins, we chose to extend these activity records by applying a method used in~\cite{sadilek2013nemesis}: for each geo-tagged tweet located within a small distance (30 meters) of a POI, we count it as a check-in from this POI. Through this process, we extend the number of check-ins to 1,028,016 for NYC and 971,660 for ROC. In order to study gender effect on lifestyles, we employ the API of genderize.io to assign gender tags to each user~\cite{abbar2014you}. Genderize.io gives the probability of an individual being either male or female given his or her username. We first filter out the users who send less than 10 tweets during our sampling period, and obtain 12,960 users in NYC and 10,576 users in ROC. We then feed the handles for these users' Twitter accounts into the genderize.io API, and filter out gender tags with low confidence (probability $< 0.8$). From this, we aggregate a total of 3,493 male- and 3,508 female-labeled users (see Table~\ref{tab:table2}). Other work has predicted users’ gender using tweet contents~\cite{schwartz2013personality}, profile information, and profile pictures~\cite{quercia2011our}; however, given the complexity of these methods, and the reasonable accuracy of our approach, we exclusively used the genderize.io API to assign gender labels. 
\begin{table}
\center
\begin{tabular}{ | l || c | c |  }
\hline
& NYC & ROC \\ \hline
 Foursquare Check-ins & 233,046 & 49,744 \\ \hline
 Foursquare venues & 99,466 & 13,483\\ \hline
 Extended Check-ins & 1,028,016 & 971,660\\ \hline \hline
 Total \# of users & 12,960 & 10,576 \\ \hline
 Male users &1,690 & 1,491 \\ \hline
 Female users & 1,803 & 2,017\\ \hline
\end{tabular}
 \caption{Number of users in our data set, by city and gender. We only assign gender labels to users when a high confidence in gender classification is achieved, so the genders of many users are considered unknown.}~\label{tab:table2}
 \vspace{-2em}
\end{table}

Two key points should be verified to ensure the quality of our data set:

\begin{figure}[!htbp]
\centering
  \includegraphics[width=0.9\columnwidth]{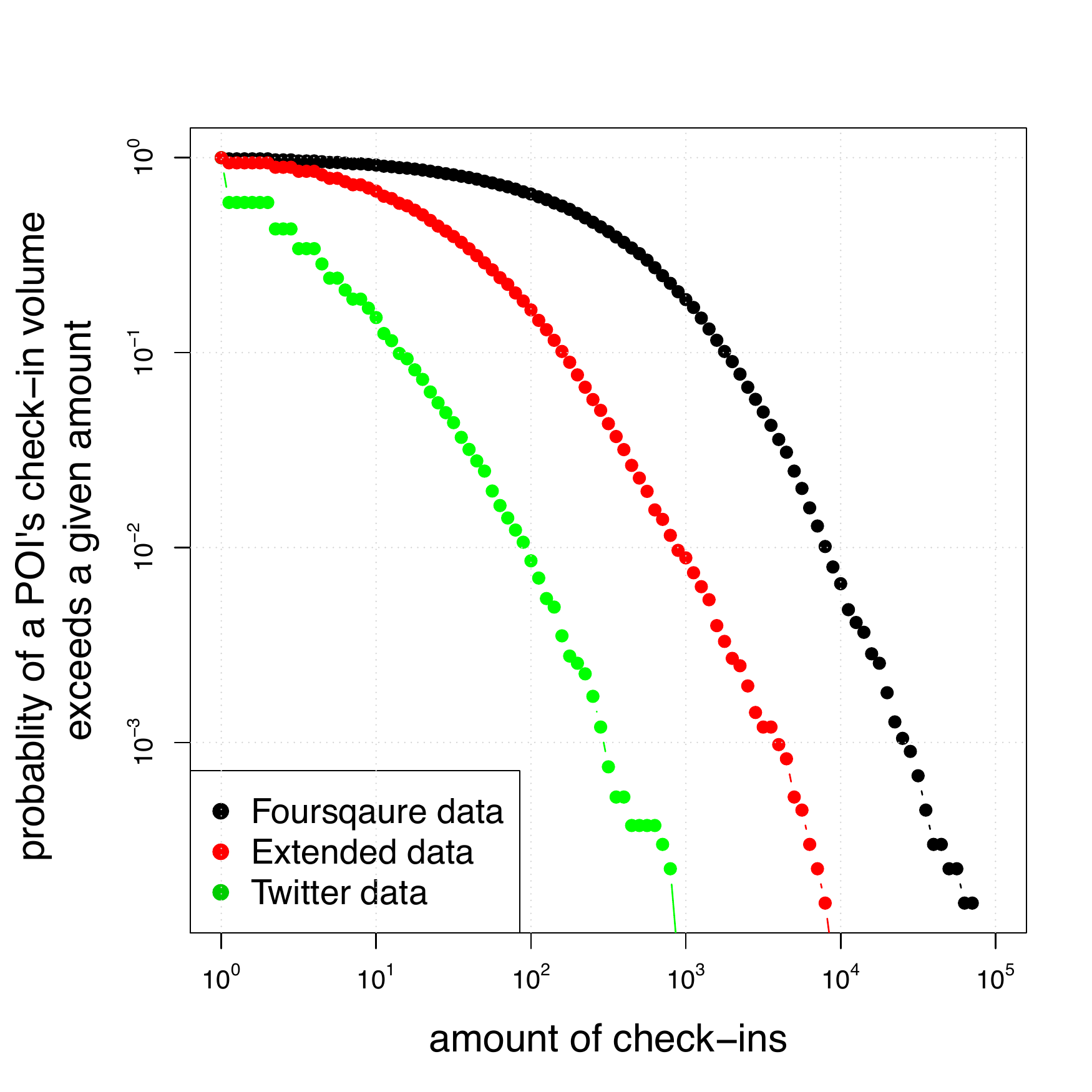}
  \caption{Plot of Complementary Cumulative Distribution Functions for the numbers of check-in amount of POIs. This plot reports the probabilities of a POI's check-in volume exceeds a given amount in three data sets. 
  }~\label{fig:long_tailed}
  \vspace{-1em}
\end{figure}

1) Only 20\% to 25\% Foursquare accounts are linked with Twitter ~\cite{foursquare_link}. Check-ins collected from tweets form a subset of all the Foursquare records. We need to ensure that our data set should have a similar distribution to the original Foursquare data. To validate the applicability of our extension method, we plot the probability of the amount of visits as a function of the amount of POIs as Noulas et al. did in~\cite{noulas2011empirical} in Fig.~\ref{fig:long_tailed}. It shows that the extended data set not only preserves the long-tailed characteristic, but also shortens the gap between the original Foursquare data and its subset that is extracted from tweets. 

\begin{figure}[!htbp]
\centering
 \includegraphics[width=1\columnwidth]{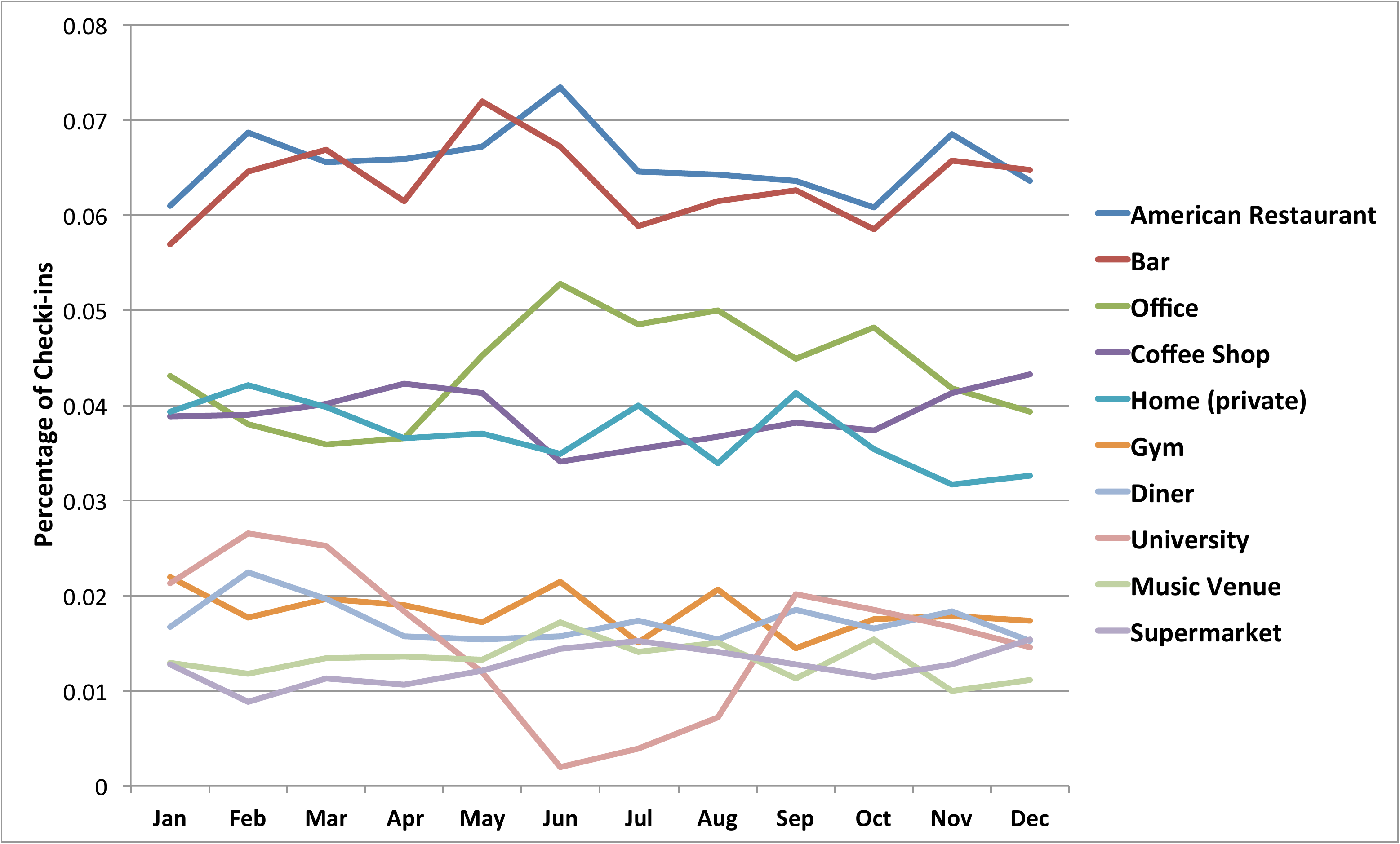}
  \caption{Percentages of check-ins from top 10 visited POI categories.}~\label{fig:12_month}
  \vspace{-1em}
\end{figure}

2) In order to obtain datasets of comparable size for the two cities, we collect tweets in NYC for a one-month period, and ROC for a one-year period. The length of the time period for tweet collection is different in NYC and ROC data sets. Tweets from NYC were posted during June 2012, while tweets from ROC were posted from July 2012 to June 2013. In Fig.~\ref{fig:12_month}, we plot the percentage of check-ins from the top 10 most frequent check-in categories. Note that we eliminate duplicate categories. For example we omit ``Restaurant'' (ranked 3rd) since we have ``American Restaurant'' in the first place. The portions of check-ins from most categories remain at stable levels throughout the year. This observation implies that the distribution in one month could approximately represent the remaining months of a year. One exception is the category of University, which shows a decrease from May to August. This coincides with the summer break of universities.

\section{Lifestyle difference at city level}

\begin{figure}[!htbp]
\centering
  \includegraphics[width=0.9\columnwidth]{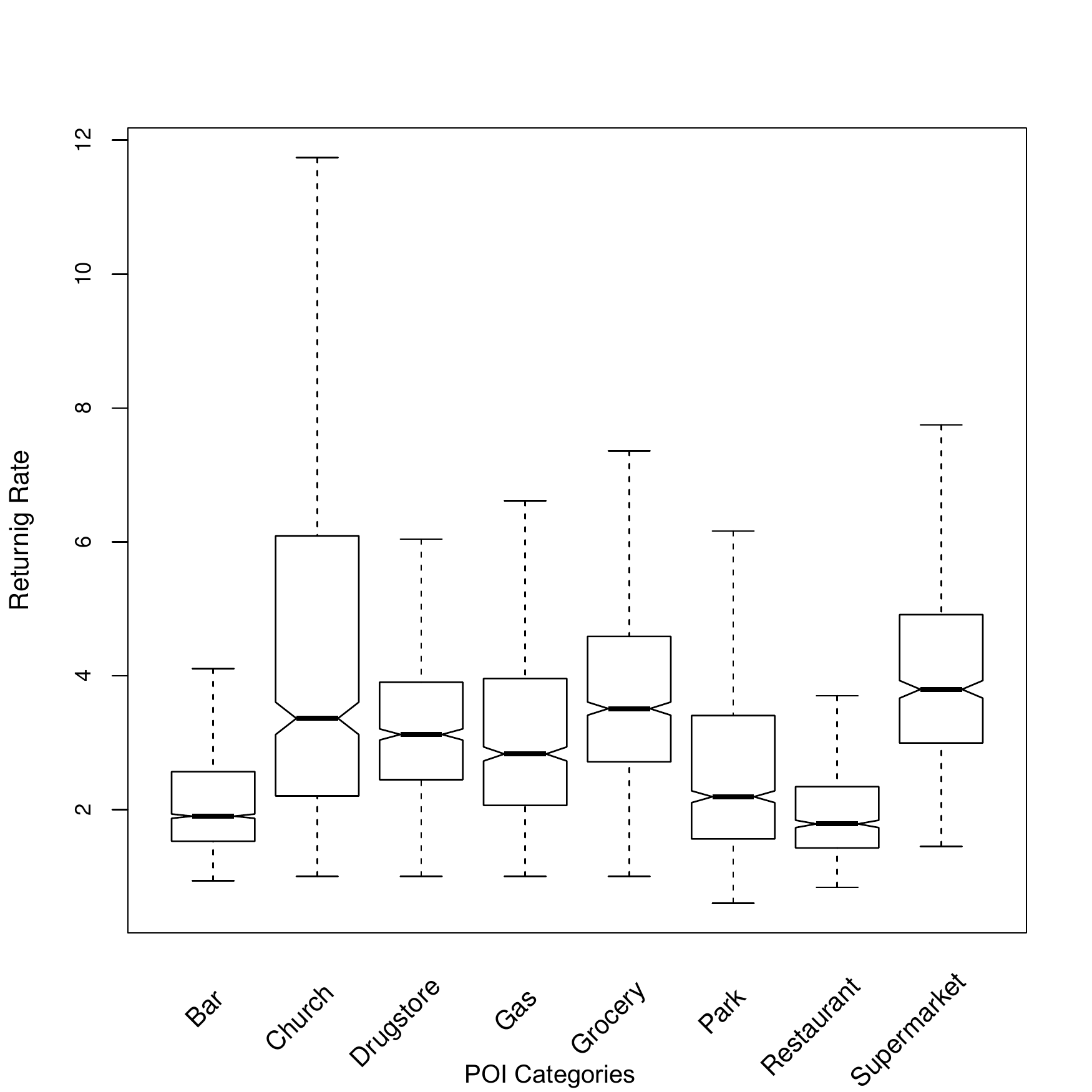}
  \caption{Box plot of visiting frequency of Bar, Church, Drugstore, Gas Station, Grocery, Park, Restaurant and Supermarket, aggregated over both cities.}~\label{fig:return_range}
\end{figure}

\begin{figure}[!htbp]
\centering
  \includegraphics[width=0.9\columnwidth]{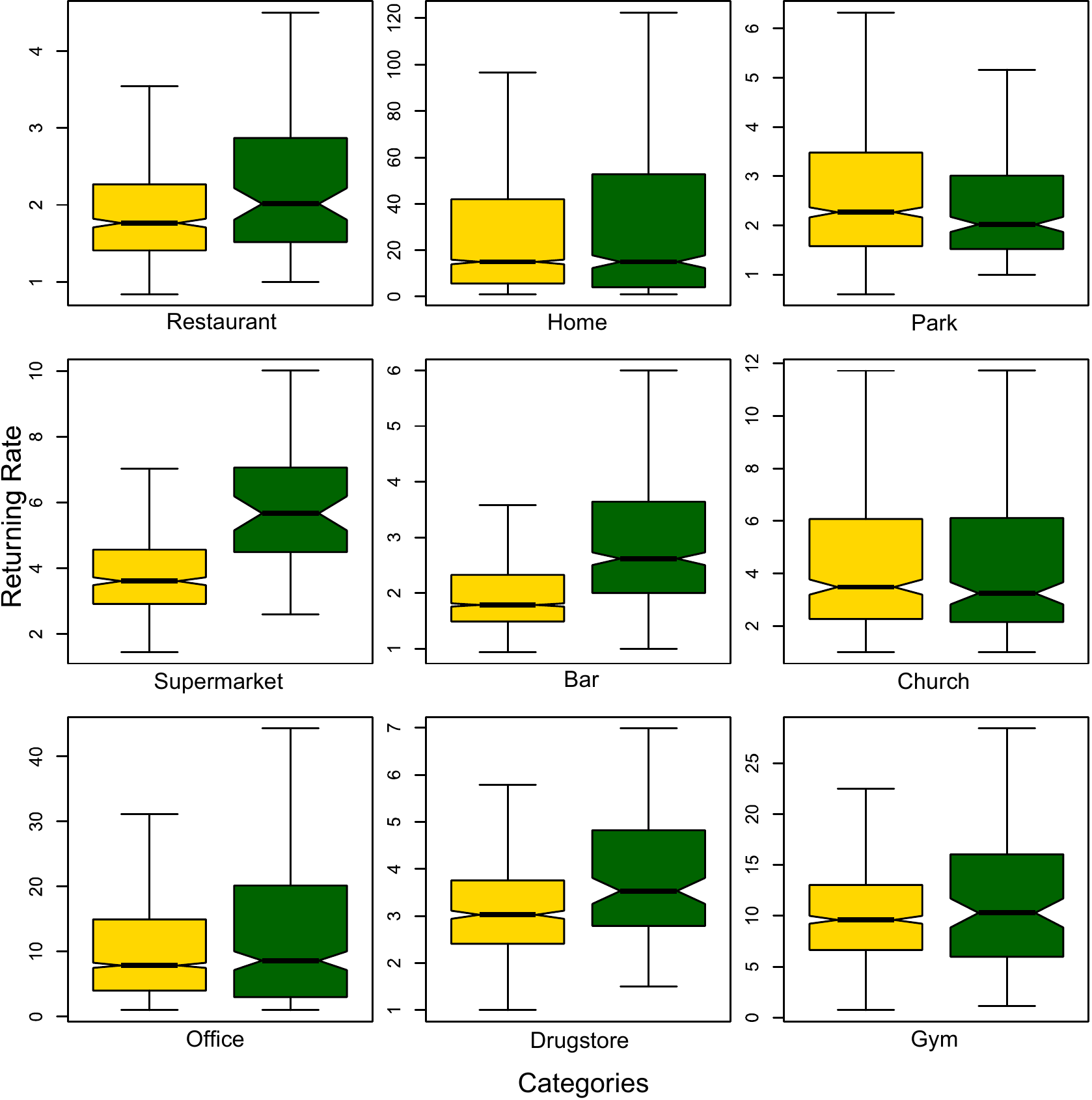}
  \caption{Comparison between visiting frequencies of 9 POI categories in big cities and small cities. The yellow boxes are the frequencies for NYC and the green ones are for ROC.}~\label{fig:return_range_compare}
\end{figure}

\subsection{Visiting Frequency of POIs}
The visiting frequency of for a location is defined as the number of visits (check-ins) divided by the number of unique visitors. In other words, visiting frequency is the average visits per visitor to a location. This metric measures the degree of relevance of a POI to people's daily life. The higher visiting frequency is, the more relevant the POI is to a person's lifestyle. ``Home'', for example, as one of the most important locations to individuals' lives, has a very high visiting frequency. Most of the check-ins at home are performed by family members or friends, so the visiting frequency of home is very high. On the contrary, a public location such as bar, usually has a lower visiting frequency.

Regarding visiting frequency, we have two interesting observations:

\begin{itemize}
\item Each POI category has a specific range of visiting frequency, which is clearly indicative of differing functionality between different POIs in people's daily life.
\item Some categories show different ranges of visiting frequency in cities with different sizes. This help us to examine the different lifestyles at the city level.
\end{itemize}

\subsubsection{Visiting frequency range of POI categories}
We plot the visiting frequency of several popular POI categories as a box plot in Figure~\ref{fig:return_range}. This plot shows that categories that are highly related to daily life are visited repeatedly, e.g. Church, Grocery, Drugstore and Supermarket all have a high visiting frequency. The median visiting frequencies of all categories is approximately 4, though some show very high visiting frequencies. For example, the visiting frequencies for some churches reaches 12, indicating a high prevalence of Church in some people's lives. As we expected, the highest visiting frequencies appear in Home (with a median of approximately 20 for both cities) and Office (with a median of approximately 10 for both cities), as shown in Figure~\ref{fig:return_range_compare} for details. These two are the most frequently visited locations for most people. The visiting frequencies of Bar and Restaurant are much lower with a median around 2.

\subsubsection{Difference in visiting frequencies between NYC and ROC}
In this section, we compare the visiting frequencies of categories in big cities and small cities (see Figure~\ref{fig:return_range_compare}). It is interesting that the visiting frequencies for some categories in small cities are larger than those in big cities, such as Restaurant, Bar, Supermarket and Drugstore. This may imply a higher regularity of life in smaller cities -- in other words, people in smaller cities are more localized, with stricter routines. In big cities, people have more options to go when eating (Restaurant), having fun (Bar) and purchasing daily necessities (Supermarket and Drugstore). Therefore, these places are generally visited less frequently in larger cities. While for other categories such as Home, Office and Church, the visiting frequencies of two types of cities are roughly the same. This makes obvious sense because the lifestyles of working, returning home and religion should be similar under the same cultural atmosphere.

\subsection{Basic mobility patterns in big cities and small cities} 
\begin{figure}[!htbp]
 
\begin{subfigure}{0.5\textwidth}
\includegraphics[width=0.9\linewidth, height=6cm]{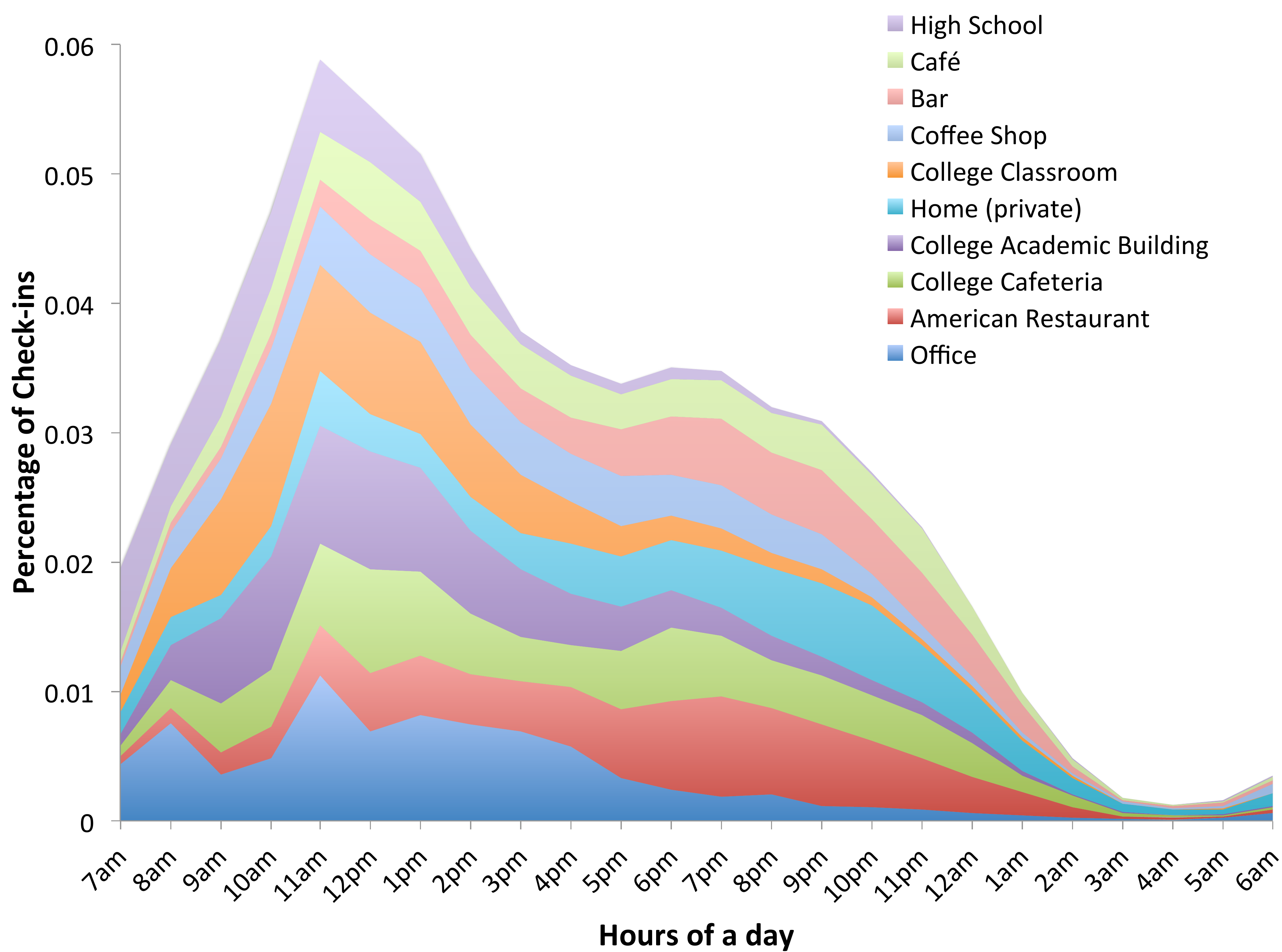} 
\caption{Rochester weekdays}
\label{fig:roc_subim1}
\end{subfigure}
\begin{subfigure}{0.5\textwidth}
\includegraphics[width=0.9\linewidth, height=6cm]{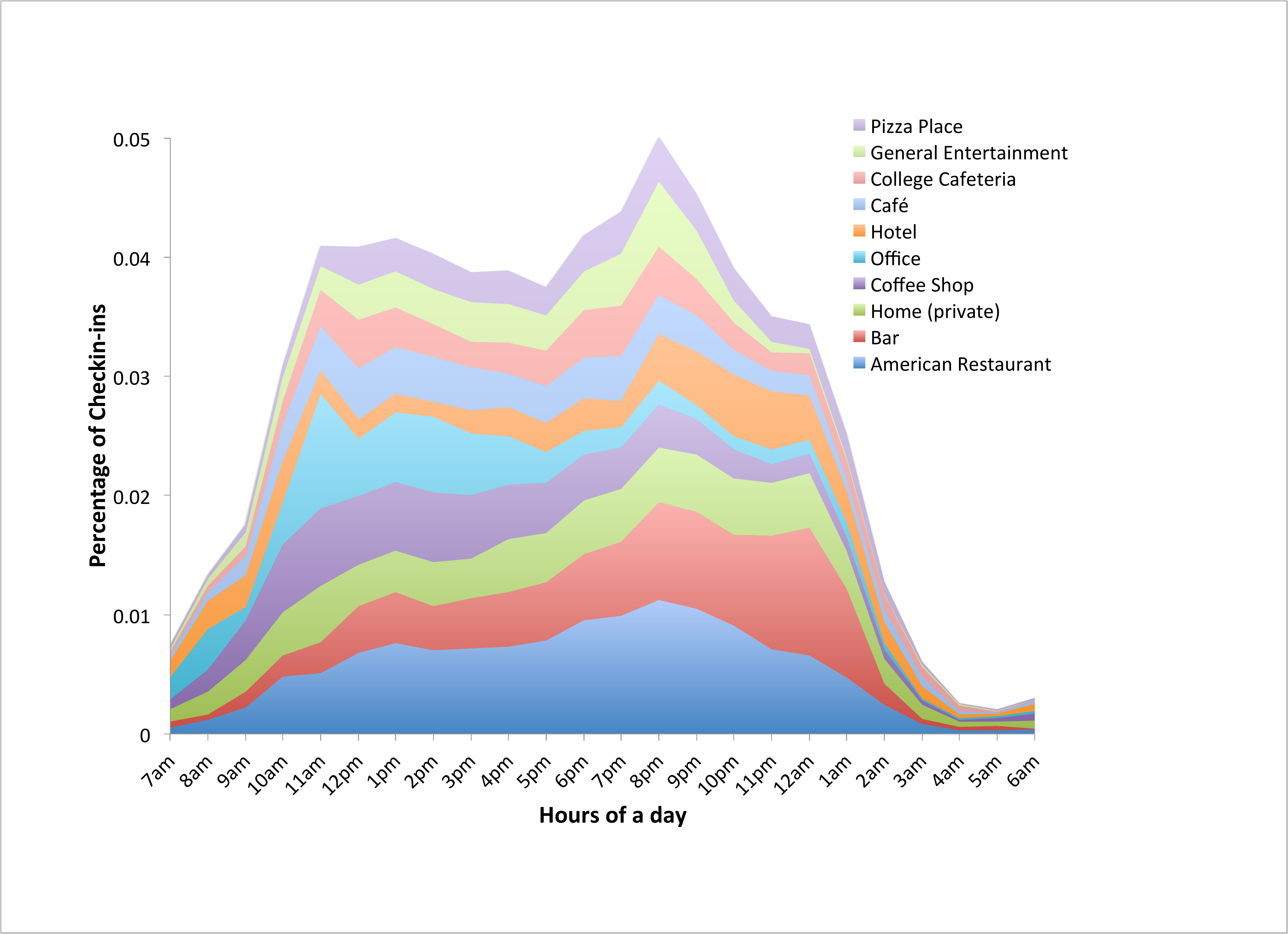}
\caption{Rochester weekends}
\label{fig:roc_subim2}
\end{subfigure}
 
\caption{Stacked plot of the 10 most popular categories over weekdays and weekends in Rochester. Categories are listed in the order of increasing probability from the top down. The width of each fault indicates the percentage of a POI category for a given time of day.}
\label{fig:roc_stacked}
\end{figure}

\begin{figure}[!htbp]
 
\begin{subfigure}{0.5\textwidth}
\includegraphics[width=0.9\linewidth, height=6cm]{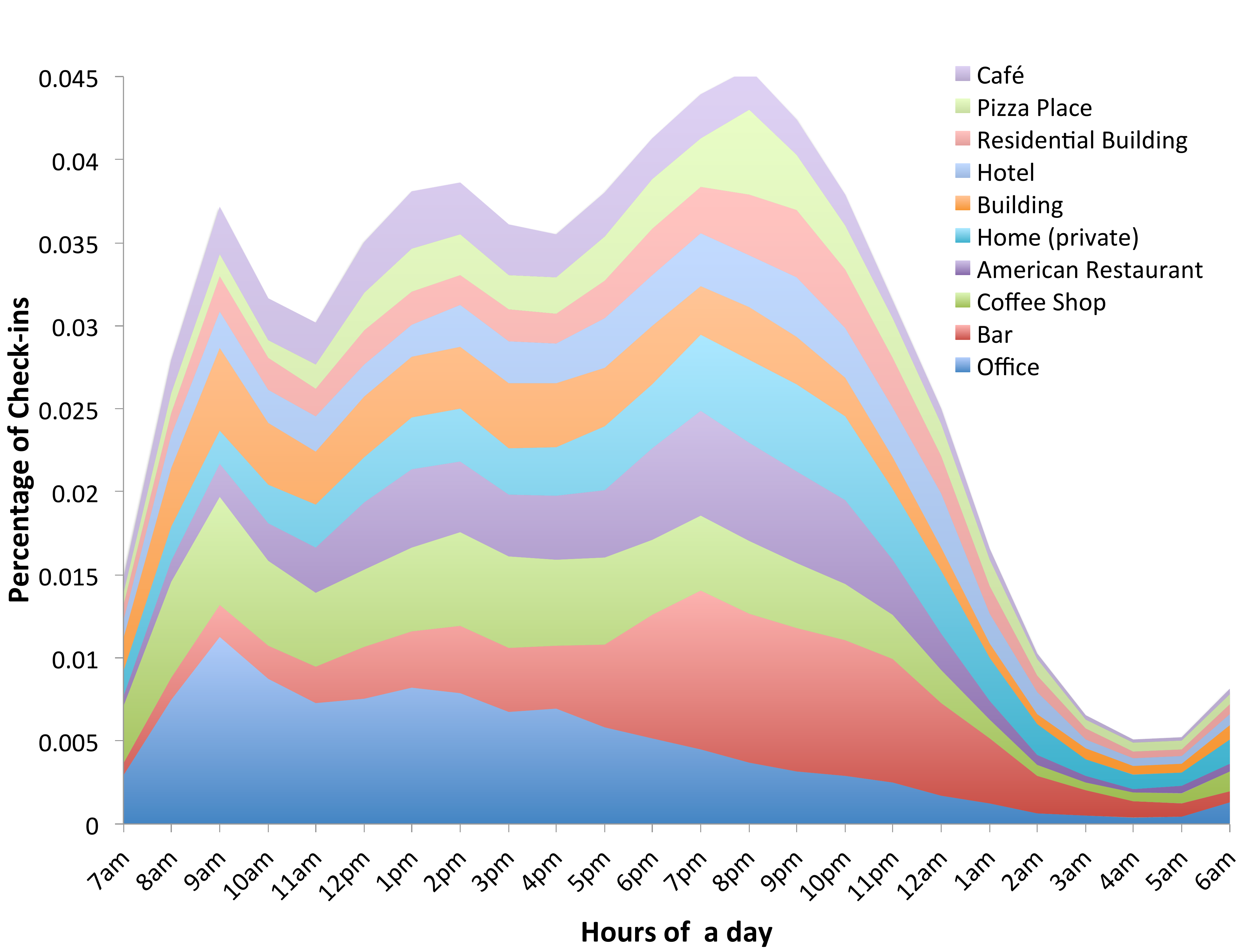} 
\caption{New York City weekdays}
\label{fig:subim1}
\end{subfigure}
\begin{subfigure}{0.5\textwidth}
\includegraphics[width=0.9\linewidth, height=6cm]{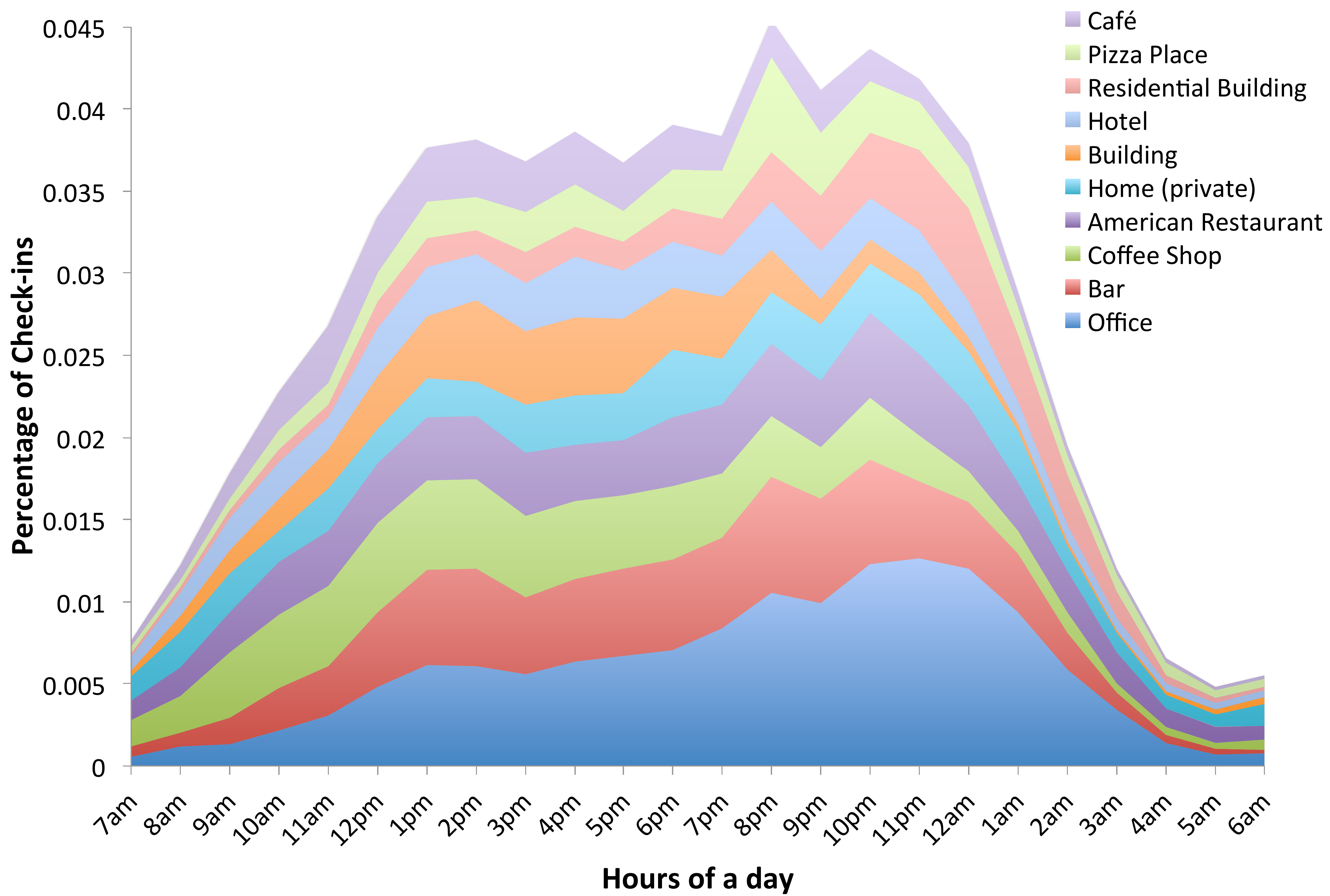}
\caption{New York City weekends}
\label{fig:subim2}
\end{subfigure}
 
\caption{Stacked plot of the 10 most popular categories over weekdays and weekends in New York City. Categories are listed in the order of increasing probability from the top down. The width of each fault indicates the percentage of a POI category for a given time of day.}
\label{fig:nyc_stacked}
\end{figure}

It is interesting to study the fluctuation of residents' activity over time in terms of occurrence at POIs. We plot the 10 most popular POI categories on weekdays and weekends separately for ROC and NYC. On weekends, the mobility patterns of the two cities are similar (Figure~\ref{fig:roc_subim2} and Figure~\ref{fig:subim2}). The total check-in amount climbs rapidly to a high level around 10am and 12pm in ROC and NYC, respectively. The activity levels remain constant until 9pm, when a peak of check-ins appears in both cities, indicating a sudden increase of mobility in weekends night. After the peak, the activity level in ROC moves down quickly, while it remains at a high level through 2am in NYC. The 3 most frequently visited POI categories on weekends for both cities are Bar, American Restaurant, Home. Obvious divergence is present between the weekday mobility patterns of the two cities (Figure~\ref{fig:roc_subim1} and Figure~\ref{fig:subim1}). In big cities, there are three peaks during a day appearing around 8am, 1pm and 9pm. Similar pattern also appear in London according to~\cite{noulas2011empirical}, indicating roughly the same mobility pattern between London and NYC. Among the three peaks, the highest one is at night, which implies that night is the most active period in large cities. In contrast, there is only one peak during a day at 10am in smaller cities and the check-in amount drops significantly after that. It reveals that during weekday nights, people in small cities are not as active as those in large cities. During the nights in weekdays, people in small cities prefer visiting Home, American Restaurant and Cafeteria, while in large cities, Bar is much more popular during night, indicating that people in large cities are more prone to indulge in copious recreation during the weekends.


\section{Mining Lifestyles with Matrix and Tensor Decomposition}

\subsection{Matrix Decomposition}

The activities of a user, $a$, can be described with an $M$ dimensional vector. For temporal patterns we set $M$ to 24 and values in the vector are the activities of the user performed in each hour, i.e. the number of check-ins. When we examine spatial patterns latent in individuals' actives, $M$  is set to be equal to the number of POI categories, and each element indicates the amount of check-ins the person performed in a single POI category. We refer to vector $a$ as an \textit{``activity vector''} of a user.

We assume that a person's activities are determined by the lifestyle(s) that person lives. Formally, $$a = w \times L$$ where L is a $k$ by $M$ matrix, recording $k$ latent lifestyles, and $w$ is a coefficient vectors of $k$ dimensions, indicating the user's preference to each lifestyle. 

To uncover and compare lifestyles that are commonly followed in different cities, first we assemble the activity vectors of residents into a single matrix for each city. We define 
$$A_{roc} = (a_{1}, a_{2}, ..., a_{N_{roc}})^T$$
where $a_{i}$ indicates the activity vector of a resident, and $N_{roc}$ is the number of samples we collected from the Great Rochester area. Similarly, 
$$A_{nyc} = (a_{1}, a_{2}, ..., a_{N_{nyc}})^T$$ 
where $N_{nyc}$ is the number of samples we collected from the Greater New York area. Second, we concatenate $A_{roc}$ and $A_{nyc}$ to obtain a complete matrix 
$$A = (A_{roc}, A_{nyc})^T$$ 
where $A$ is a $(N_{roc} + N_{nyc})$ by $M$ matrix.

Third, we decompose $A$ into two matrix $W$ and $L$. $W$ is a $N$ by $k$ coefficient matrix, while $L$ is the lifestyle matrix we explained above. Since non-negative matrix factorization (NMF) usually leads to interpretable results~\cite{lee1999learning}, we applied it to complete the decomposition. Formally, we solve the following optimization problem:

$$ \underset{W,L}{min} \frac{1}{2}\|A- WL\|^2_F \quad s.t. \quad L\geq 0, W\geq 0 $$
where $A\in\mathbb{R}^{(N_{roc} + N_{nyc})\times M}$, $W\in\mathbb{R}^{(N_{roc} + N_{nyc})\times k}$, $L\in\mathbb{R}^{k\times M}$. $\|X\|_F =(\sum_{i,j}|X_{ij}|^2)^{-\frac{1}{2}}$ is the Frobenius norm, $L \geq 0$ (or $W \geq 0$) requires that all components of $L$ (or $W$) should be nonnegative. $L$ uncovers the lifestyles that people follow, while $W$ provides information about individuals' preferences across these lifestyles. 

\begin{figure}[!htbp]
\centering
  \includegraphics[width=0.9\columnwidth]{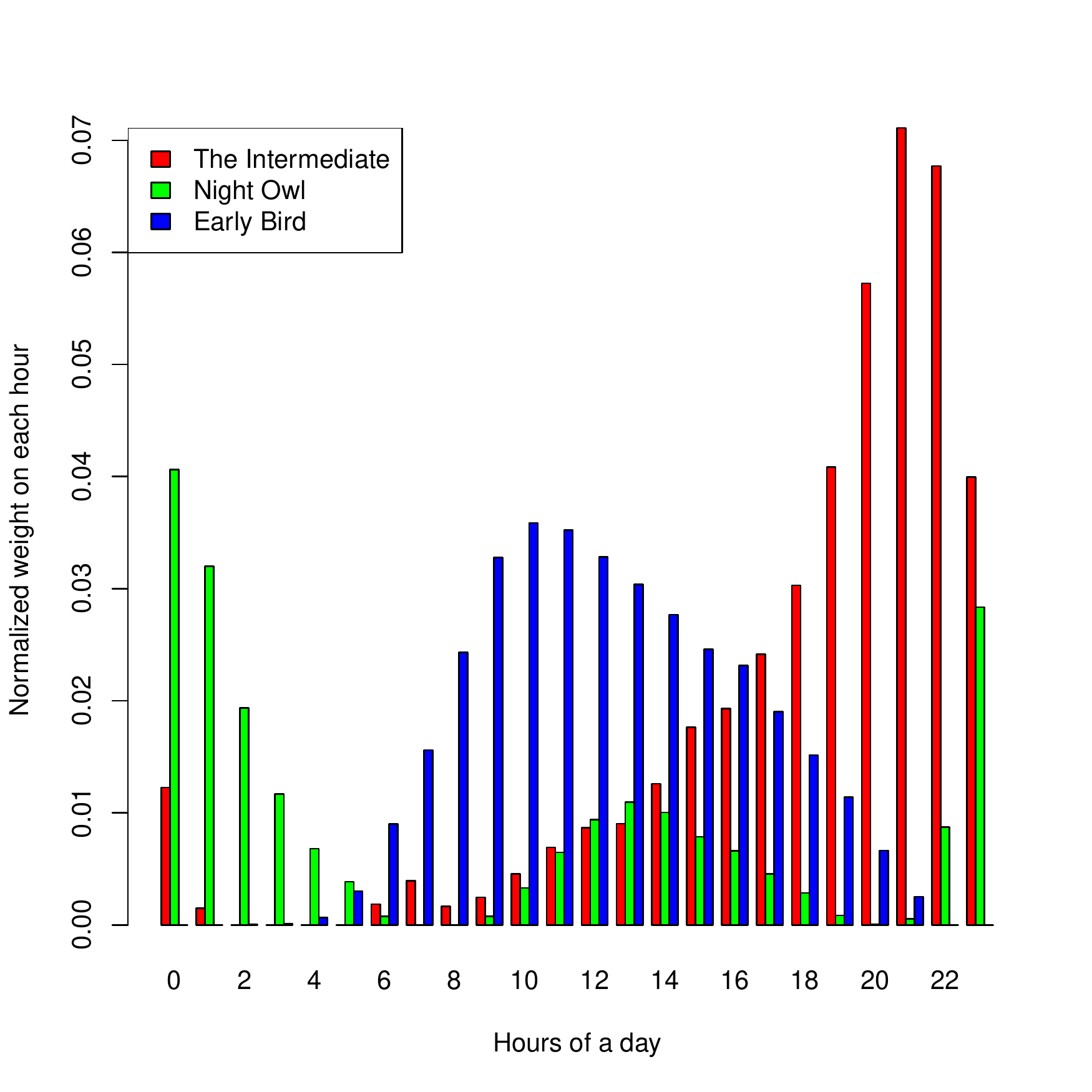}
  \caption{Active time ranges of night owls, early birds and intermediates over weekdays.}~\label{fig:time_weekday}
  \vspace{-1.5em}
\end{figure}

After decomposition, we split $W$ into smaller matrices, each of which records a sample of lifestyles for various groups of people. On a city level, $W$ is split into two smaller matrices, $W = (W_{roc}, W_{nyc})^T$, where $W_{roc}\in\mathbb{R}^{N_{roc}\times k}$ and $W_{nyc}\in\mathbb{R}^{N_{nyc}\times k}$. At a finer granularity, $W$ consists of four even smaller matrices:  $W = (W_{roc \ male}, W_{roc \ female}, W_{nyc \ male}, W_{nyc \ female})^T$. For a particular group of people, i.e. a component matrix, the degree of preference for a lifestyle is defined as the average of the coefficients of people in the group for the lifestyle. For example, the preference to $i_{th}$ lifestyle of residents of New York City is calculated by averaging the $i_{th}$ column of matrix $W_{nyc}$.

In the following sections, we report the temporal and spatial lifestyles found from people's activities, and compare the preferences to lifestyles in the two cities.

\subsection{Third-Order Tensor Decomposition}

\begin{figure}[!htbp]
\centering
  \includegraphics[width=1.0\columnwidth]{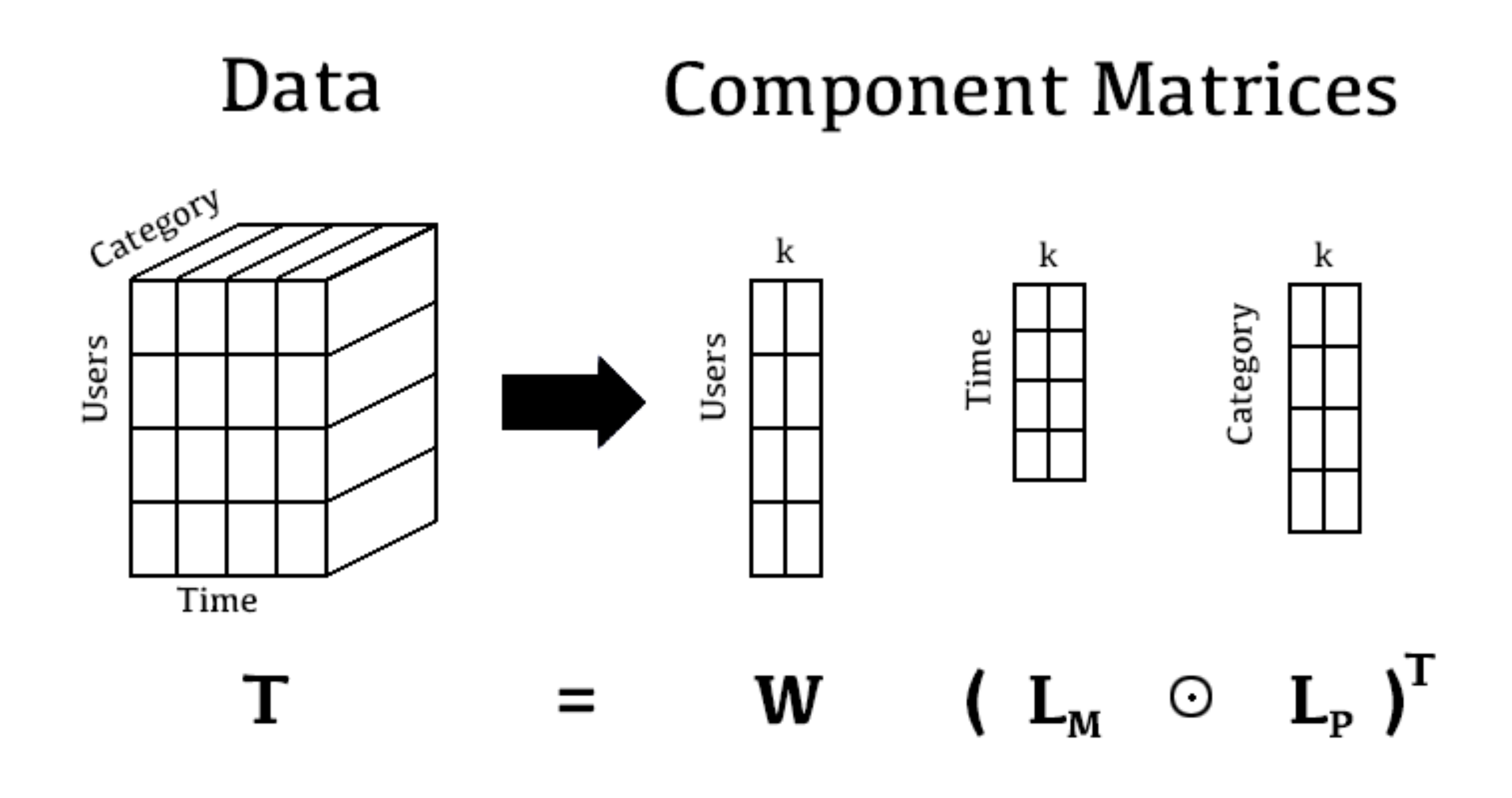}
  \caption{Visualization of tensor cube $T$, decomposed into component matrices $W$, $L_M$, and $L_P$.}~\label{fig:tensor-viz}
\end{figure}

User activities may be analyzed across multiple dimensions simultaneously using higher-order tensors. Tensors are a natural way to aggregate data across multiple factors. Vectors and matrices are special cases of tensors, where each vector $v$ of dimensionality $D$, it is true that $v \in \mathbb{R}^{D}$, and for each matrix $M$ of dimensionality $D_1$ by $D_2$, $M \in \mathbb{R}^{D_1 \times D_2}$. Tensors generalize this to data structures of arbitrary order, where vectors are of order 1, matrices of order 2. For tensor $T$ of order $d$, $T$ may be concisely described as: $T \in \mathbb{R}^{D_1 \times D_2 \times \ldots \times D_d}$. To learn temporal-spatial patterns of human activities, for example, we can aggregate the data into a third-order tensor. In such a tensor, a person's activities is recorded as a matrix, of which dimensions are POI categories and hours of a day. Decomposition on the tensor produces multidimensional knowledge on lifestyles. In Fig.~\ref{fig:tensor-viz}, we illustrate the tensor decompositio processes.


The most commonly used technique for tensor decomposition is CANDECOMP/PARAFAC (CP) decomposition~\cite{kolda2009tensor}. This algorithm decomposes a tensor of order $d$ into $d$ separate matrices, each of dimensionality $k \times D_j$, where $k$ is the number of components selected a priori and $D_j$ is the dimension for the tensor's $j^{th}$ order.

The formal optimization problem for this decomposition is: $$ \underset{W,L_M,L_P}{min} \|T - W (L_M \odot L_P)^\top \|$$ 

In this equation, $\odot$ represents the Khatri-Rao product. The Khatri-Rao product may be considered as a column-wise Kronecker product $\otimes$ between two matrices with equal numbers of columns $A = [a_1, a_2, a_3]$ and $B = [b_1, b_2, b_3]$, where $A \odot B = [a_1 \otimes b_1, a_2 \otimes b_2, a_3 \otimes b_3]$. While $A$ and $B$ both have 3 columns here, this can be generalized for any number of columns. Assuming that $A \in \mathbb{R}^{M \times K}$ and $D \in \mathbb{R}^{N \times K}$, the Khatri-Rao product matrix will be of dimensionality $MN \times K$.

To solve the optimization problem for CP decomposition, we use the alternating least-squares (ALS) algorithm, originally proposed by ~\cite{harshman1970foundations, carroll1970analysis}. The specific implementation of CP-ALS used is provided in the \textit{scikit-tensor toolkit}. \footnote{https://github.com/mnick/scikit-tensor} At a high level, ALS incrementally uses $W$ and $L_M$ to estimate $L_P$, then $L_M$ and $L_P$ to estimate $W$, and so on, improving estimations of one matrix in each iteration.

As ALS monotonically decreases error rate for the optimization, the algorithm is subject to getting trapped in local minima. Thus ALS is not guaranteed to find an optimal solution, and results may depend heavily on initialization. In our experience, it was found that using both singular vector and random initializations converged to similar decompositions with similar error rates when using a termination condition of $10^{-5}$ error improvement between iterations.


Similar to our matrix decomposition methodology, we assume that individuals' check-in activity may be decomposed into a weighted combination of lifestyle factors stored in a matrix: $$t = w\ (L_M \odot L_P)^\top$$ where $L_M \in\mathbb{R}^{k \times M}$, and $L_P \in \mathbb{R}^{k \times P}$, each recording $k$ latent lifestyles, where again $w \in \mathbb{R}^{k}$ is a coefficient vector for a single user. $L_M$ reveals the first dimensional characterizations of each lifestyle component, and $L_P$ the second dimensional characteristics. As with our matrix decomposition framework, we consider weight matrix $W$ as a concatenation of four smaller matrices according to city and gender. However, in the work presented here, we found no significant differences in mean component weights across these demographics.

Full tensors $T_i=\{t_1, t_2, \ldots, t_N\}$ that we consider concatenate lifestyle matrices $t$ across all users. In this work, we present an analysis of two third-order tensors $T_1$ and $T_2$, such that $T_i \in \mathbb{R}^{N \times M \times P}$. $N$ indexes check-in counts by user id and $P$ indexes by category. Only the 100 categories with the highest number of check-ins are used, so $P=100$ for both $T_1$ and $T_2$. $T_1$ indexes by times of day as well, so $M=24$. $T_2$ indexes instead by days of the week, so $M=7$. 

The first tensor, $T_1\in\mathbb{R}^{N \times 24 \times 100}$, will allow us to examine joint spatial-temporal lifestyles, indicative of user's locational behavior at various times of the day. Trivially, we might find components of user check-in at bars and pubs, with greater weight assigned to night hours than to the morning or afternoon. The second tensor, $T_2\in\mathbb{R}^{N \times 7 \times 100}$, will be conducive to locational lifestyles with distinct trends across the work week, through the weekend. For example, we might see lifestyles of individuals visiting restaurants and entertainment venues later in the week, with less weight assigned to Monday, Tuesday, and Wednesday.

\section{Temporal Aspects of Lifestyle}

\begin{figure}[!htbp]
\centering
  \includegraphics[width=0.9\columnwidth]{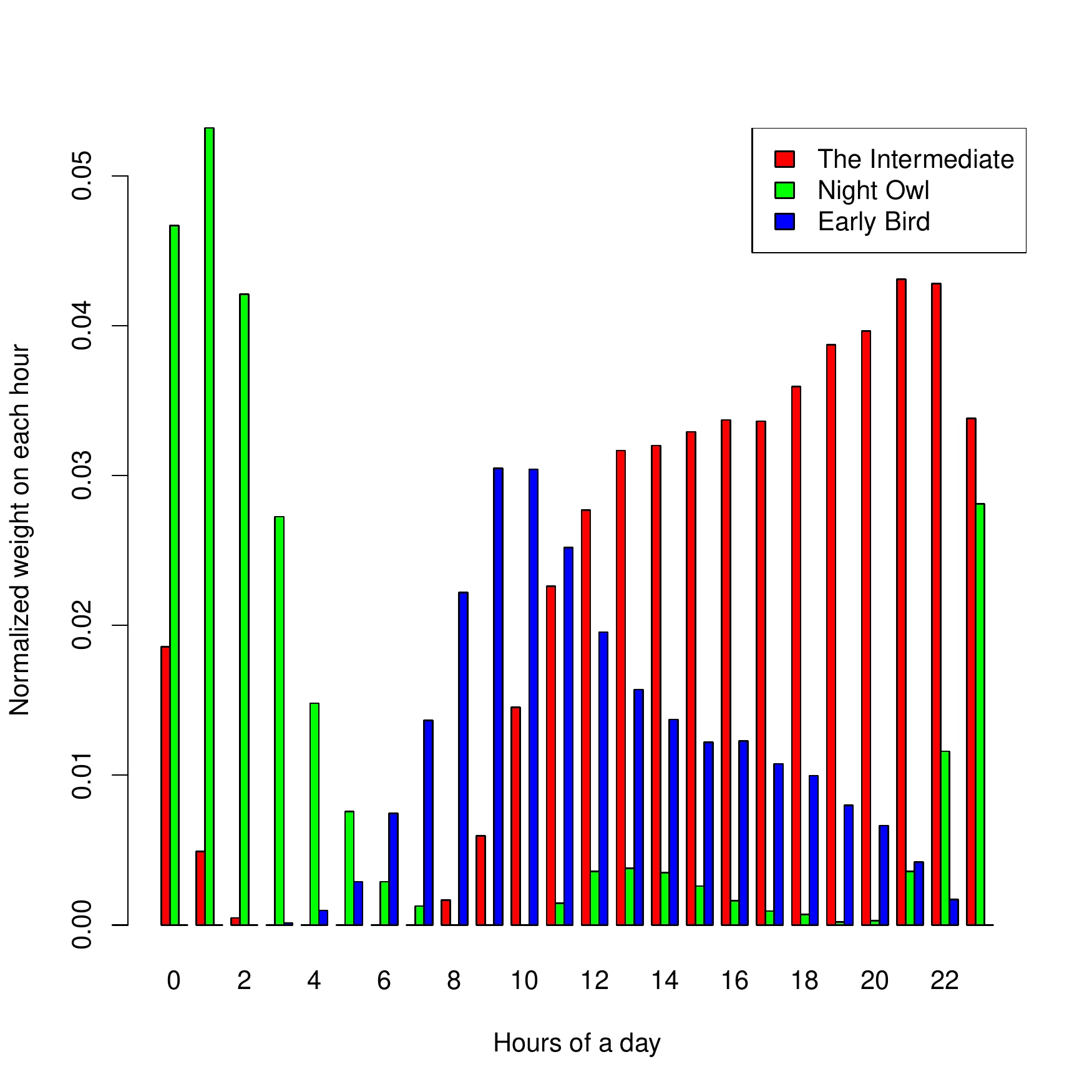}
  \caption{Active time ranges of night owls, early birds and intermediates over weekends.}~\label{fig:time_weekend}
  \vspace{-1.5em}
\end{figure}

Analogous to ETs, MTs, and NTs in the circadian topology literature, we classify people's work and rest habits into three categories:  \textit{night owls}, people who tend to stay up until late at night; \textit{early birds}, people who usually get up early in the morning and go to bed early in the evening; and \textit{intermediates}, people who have schedules between night owls and early birds~\cite{horne1975self}. 

Interestingly but not surprisingly, our approach provides support for these three common temporal lifestyles. Moreover, we are able to provide a precise description of activity level along time of day for each lifestyle. We study weekday and weekend separately to gain a better understanding of people's lives.

Let $A_{weekday}$ be a $(N_{roc} + N_{nyc})$ by $M$ matrix, where $M$ equals to 24. A component $a_{ij}$ in the matrix denotes the $i_{th}$ user's total number of check-ins during $j_{th}$ hour of weekdays. Similarly, $A_{weekend}$, also a $(N_{roc} + N_{nyc})$ by $M$ matrix, records the activities of users on weekends.  We set $k$ as 3 to align with the number of predefined categories, and then employ matrix decomposition on $A_{weekday}$ and $A_{weekend}$, respectively. The results are $L_{weekday}$ and $W_{weekday}$ for $A_{weekday}$;  $L_{weekend}$ and $W_{weekend}$ for $A_{weekend}$. We first plot the result matrices $L_{weekday}$ and $L_{weekend}$ in Fig.~\ref{fig:time_weekday} and Fig.~\ref{fig:time_weekend}.

\begin{figure}[!htbp]
\centering
  \includegraphics[width=0.9\columnwidth]{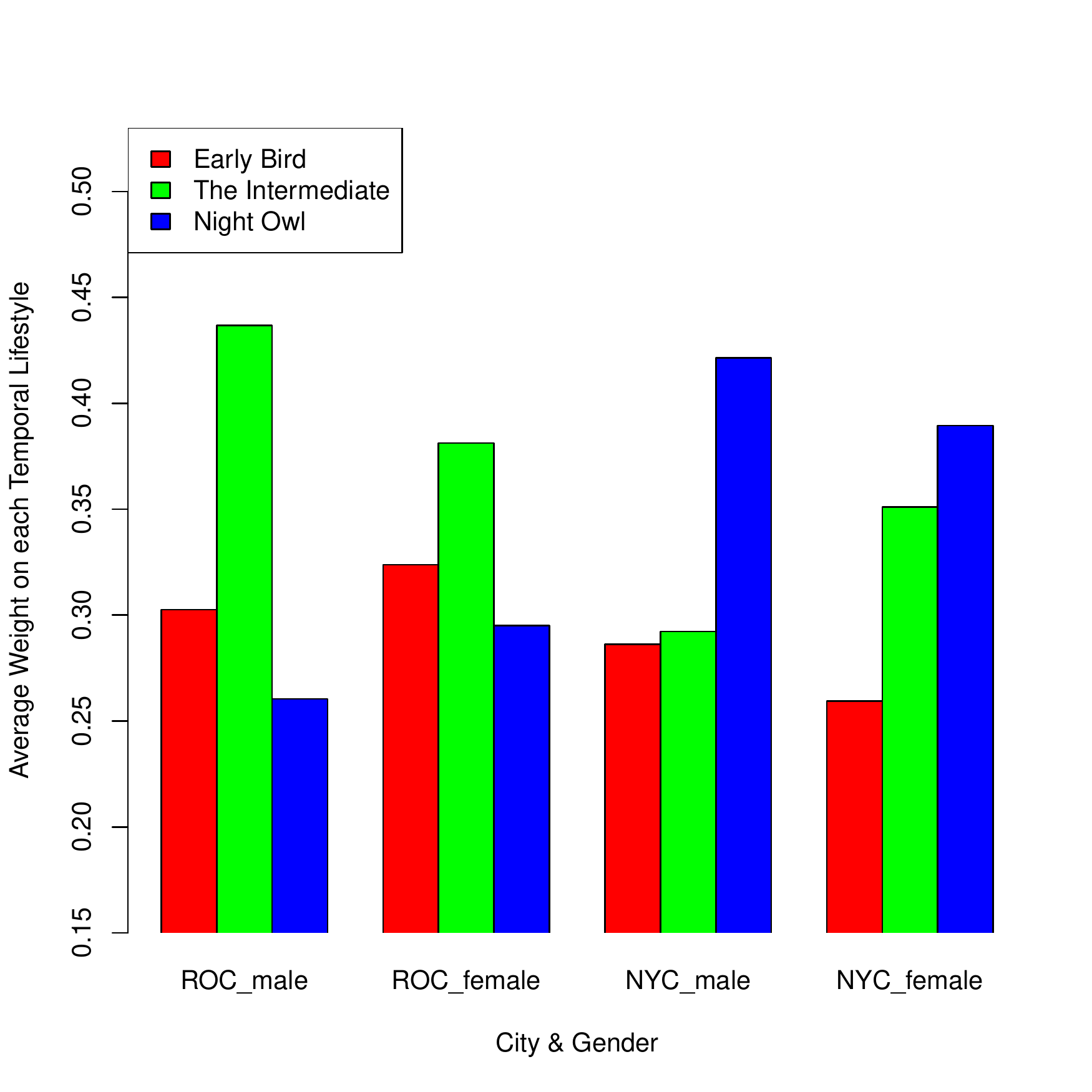}
  \caption{Average weight on night owls, early birds and intermediates of male and female residents of a small and big city over weekends.}~\label{fig:habit_weekend}
\end{figure}

\textit{Early birds}:
In weekdays, early birds' days start around 7 am, and they are most active around noon. After that, their activities decrease, and then vanish gradually in the night around 10 pm. The distribution of activity over time in weekend is similar for early birds, except the increase and decrease of activities in weekend is faster, which leads to a sharper peak around 12 pm. 

\textit{Night Owls}:
For night owls, we observe two active periods in a day. Their activities start from 10 am for both weekdays and weekends. We observe the first small peak appears at 2 pm, but is not comparable to their active level during night. After the inactive daytime, the activities of night owl rocket from 10 pm and achieve maximum at late night (1 am in weekdays and 2 am weekends). Their activities vanish in the early morning (6 am).

\textit{Intermediates}:
People who are neither early birds nor night owls usually start their day in the late morning (10 am). On weekdays, the active level increases gradually in the afternoon and reaches the peak around 10 pm, and rapidly decreases to zero around 2 am. On weekends, they are more active during afternoon than on weekdays. Instead of a gradual increase, the active level grows faster after being activated (11 am) and remains high till the peak of night (11 pm).

The results extracted through data agree with the time ranges defined by traditional studies. We compare our results with those from previous human Circadian Rhythm ~\cite{shahid2012morningness}.

In table~\ref{tab:table3} we list the time ranges of wake up time, sleep time and most active time of the corresponding types (morning-type, evening-type, neither-type) in the morningness-eveningness questionnaire (MEQ)~\cite{horne1976self}, to compare with these time ranges we list the percentage of activity during the same time range of each lifestyle decomposed from our data using NMF. We define these three time ranges as: from the early morning (5 am), the time range of the first $\sim 15\%$  of activities is defined as ``get up'', the next $\sim 70\%$ is defined as ``most activity'' and the final $\sim 15\%$ is defined as ``go to bed''. These percentages are not exact, and a small amount of activity is present between the ``go to bed'' and ``wake up'' time ranges.  For most time ranges, our results generally agree with those from previous work, e.g. ``get up'' time range, and ``go to bed'' time range for the Early Bird and Intermediate lifestyle.

All the ``most active'' ranges in our findings are later than the previous assessments, and the ``go to bed'' for Night Owl is much later than that of the evening-type in previous work. Our explanation for these differences is twofold: Firstly, we believe that as a general trend people's activities shift a lot into night in modern times when compared with the year when the seminal previous work was done (1976) \cite{horne1975self}. Secondly, individuals' behaviors in our model are modeled by a weighted combination of ``lifestyles'', whereas in the MEQ paradigm, an individual is assigned to a single, discrete ``type''. As a consequence, our ``lifestyle'' patterns should capture more distinctive work-rest activities a single individual might follow as a subset of all his or her behaviors, whereas each MEQ ``type'' should capture the aggregate work-rest patterns for all of an individual's behaviors. For example, an individual in our model might be a ``night owl'' on the weekends and an ``early bird'' on the weekdays, while in the MEQ model this individual would be considered as either a morning-type or an evening-type.

\begin{table}

\center
\begin{tabular}{|l||c|c|}
\hline
 & MEQ & Our results\\
\hline
Early Bird &  \multicolumn{2}{c|}{} \\
\hline
Get up& 5:00 am - 7:45 am& 6:00 am - 8:00 am\\
Most active& 5:00 am - 10:00 am& 7:00 am - 2:00 pm\\
Go to bed& 8:00 pm - 10:15 pm& 8:00 pm - 10:00 am\\
\hline
Inter & \multicolumn{2}{c|}{}  \\
\hline
Get up&7:45 am - 9:45 am & 8:00 am - 10:00 am\\
Most active& 10:00 am - 5:00 pm& 2:00 pm - 8:00 pm\\
Go to bed& 10:00 pm - 12:30 am& 10:00 pm - 12:00 am\\
\hline
Night Owl& \multicolumn{2}{c|}{}   \\
\hline
Get up& 9:45 am - 12 pm & 10:00 am - 12:00 pm\\
Most active& 5:00 pm - 5:00 am& 9:00 pm - 1:00 am\\
Go to bed& 12:00 am - 3:00 am& 3:00 am - 6:00 am\\
\hline
\end{tabular}
\caption{Comparison between our method with traditional methods}
~\label{tab:table3}
\end{table}

\begin{figure}[!htbp]
\centering
 \includegraphics[width=0.9\columnwidth]{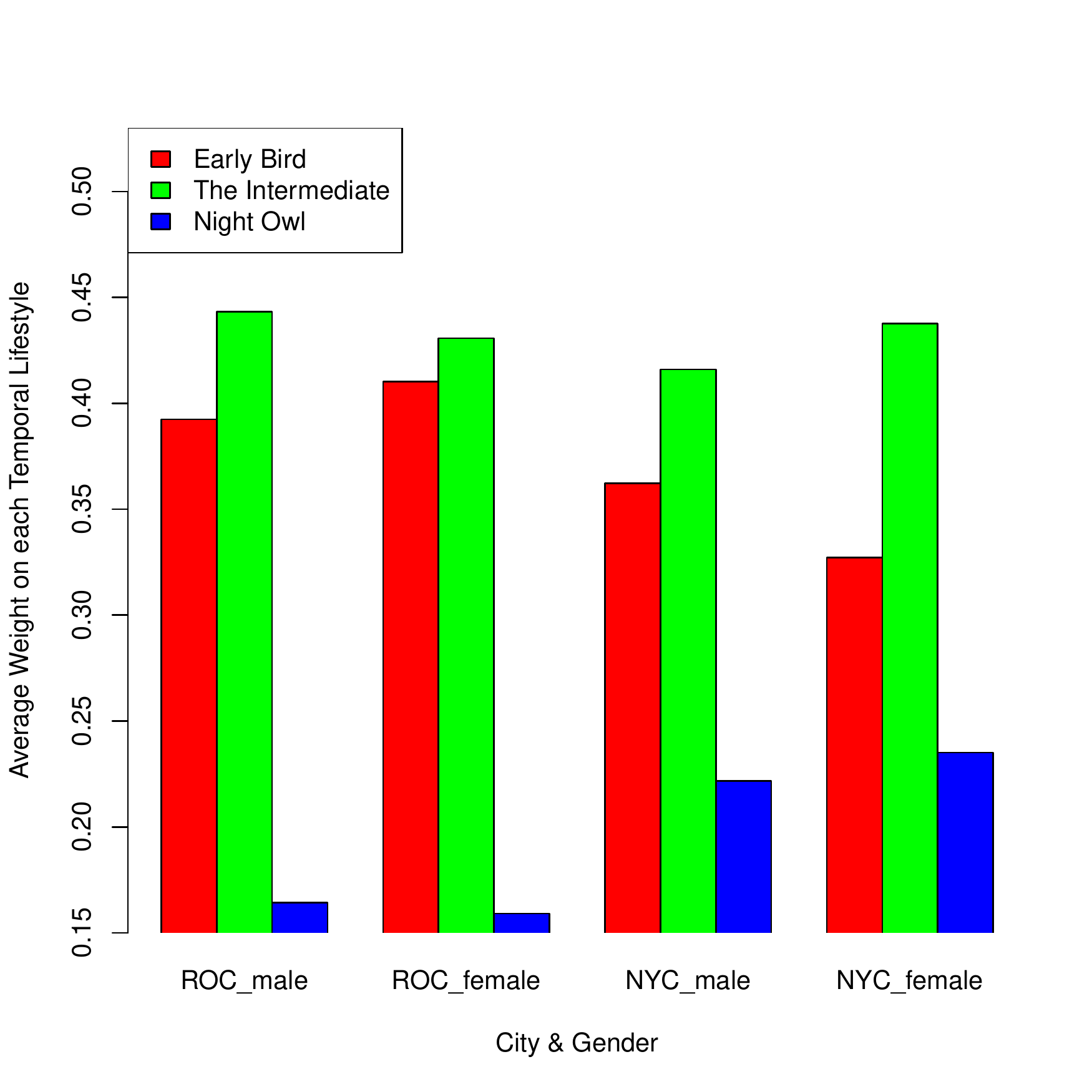}
  \caption{Average weight on night owls, early birds and intermediates of male and female residents of a small and big city over weekdays.}~\label{fig:habit_weekday}
\end{figure}

$W_{weekday}$ and $W_{weekend}$ indicate the preference of each user to three lifestyles on weekday and weekend, respectively. For each matrix, we first split it to 4 smaller matrices according to user's city (ROC or NYC) and gender. Secondly, we calculated the average preference of each group to each lifestyle. We plot the results in Fig.~\ref{fig:habit_weekend} and Fig.~\ref{fig:habit_weekday}.

Generally speaking, the average preference to night owl lifestyle on weekends is significantly higher than on weekdays. Correspondingly, the average weight of early bird and the intermediate style is significantly lower on weekends. This indicates that people in both cities are more willing to stay active late on weekends, while getting up early on weekdays. People live in big cities usually have higher preference to the night owl type for both males and females, while people tend to have higher weights on the early bird type in small cities. This observation suggests that big cities are more active than small cities during night. We did not observe significant difference between two genders on their preference to temporal lifestyles in both two cities. This agrees with previous work~\cite{merritt2003gender}, in which the authors verify that daily activity levels generally are not biased by gender. 

\subsection{Spatial Aspects of Lifestyles}
Individuals’ preferences towards specific locations are another important indicator to their lifestyles. These lifestyles can be described as combinations of several specific POI categories. For example, we observed the co-occurrence of Home, Grocery and Gas Station in many people's visiting records; we could have the feeling that the people performed this pattern are ``home-originated'', since the places they visited are quite related to daily life. Another commonly observed combination is Bar, Pub and Music Venue; people following this pattern clearly tend to have much excitement (alcohol) in their daily life. These movement patterns are conducive to understanding individuals' lifestyle preferences. 

\begin{table*}
\centering
\begin{tabular}{ | l || c | c | c | c | c | }
\hline
Hidden Patterns & 1st category & 2nd category & 3rd category & 4th category & 5th category\\ \hline
1,  College  & Residence Hall & Co-working Space & Lab & Rec Center & Wine Bar\\ \hline
2,  Restaurant & American Restaurant & Grocery Store & Supermarket & Fast Food & Diner\\ \hline
3,  Bar \& Pub & Bar & Music Venue & nightclub & Lounge & Rock Club\\ \hline
4,  Office & Office & Co-working Space & Building & Conf. Room & Bar\\ \hline
5,  Home \& Grocery & Home (private) & Supermarket & Grocery Store & Drugstore & Church\\ \hline
6,  Entertainment & Arts \& Entertainment & Baseball Stadium & Bar & Burger Joint & Concert Hall\\ \hline
7,  Sports & Gym & Yoga Studio & Athletics \& Sports & Spa & Fitness Center\\ \hline
8, Park \& Outdoor & Park & Neighborhood & Scenic Lookout & Plaza & Beach\\ \hline
9, Hotel \& Bar & Hotel & Lounge & Cocktail Bar & Roof Deck & Airport\\ \hline
10, Commute & Train Station & Subway & Train & Platform & Bus Station\\ \hline
\end{tabular}
 \caption{15 Hidden patterns with their assigned names and top 5 weighted categories of POIs of each pattern.}~\label{tab:table1}
 \vspace{-1em}
\end{table*}

We also employ the NMF method to detect these hidden patterns. Instead of a temporal activity matrix, we decompose a spatial activity matrix $A$ in this case. $A$ is a $(N_{roc} + N_{nyc})$ by $M$ matrix, where $M$ is the number of categories of POIs. The decomposition generates two result matrices $L$ and $W$. $L$ records the spatial lifestyles that are lived by the people in different cities, $W$ contains the information of the preference of each resident to these spatial lifestyles. $k$ is empirically set to 5 to achieve a good tradeoff between granularity and interpretability. 

We report the lifestyles extracted from the data in Table~\ref{tab:table1}. For each pattern we list the top 5 weighted categories of POIs and assign a name to the pattern. We can sense the clear connection between the POI categories within a hidden pattern. Take pattern one as an example. The top three weighted categories are College Residence Hall, Co-working Space and College Lab. This is a common mobility pattern of college students. Pattern seven describes a pattern of people who like to exercise, where the top three categories are Gym, Yoga Studio and Athletics \& Sports. For pattern ten, the top three weighted categories are Train Station, Subway and Train. This is a typical movement pattern of people who commute a lot. 

It's natural to see one's behaviors as a combination of several lifestyles. For example, for a college student we may find he/her lives first lifestyle in Table~\ref{tab:table1} (College lifestyle) as well as the third (Bar \& Pub lifestyle) and the seventh (Sport lifestyle) with different weights. By ``weight'' we mean the importance of a certain lifestyle to one's daily life. Rows in $W$, indicated by $w_{i}$, are such weight vectors. $w_{i}$ is a 15 dimensional vector, indicating the quantified preferences of $i_{th}$ user to 15 spatial lifestyles. In order to gain a group level understanding of lifestyles, we employ a clustering method on $w_{i}$s. The center of each cluster denotes the mean lifestyle combination for a group of individuals. Moreover, by analyzing the component of a group we are able to determine the tastes for lives of residents cities of different size. We set the number of clusters to 5 empirically.

\begin{figure}[!htbp]
\centering
  \includegraphics[width=1.0\columnwidth]{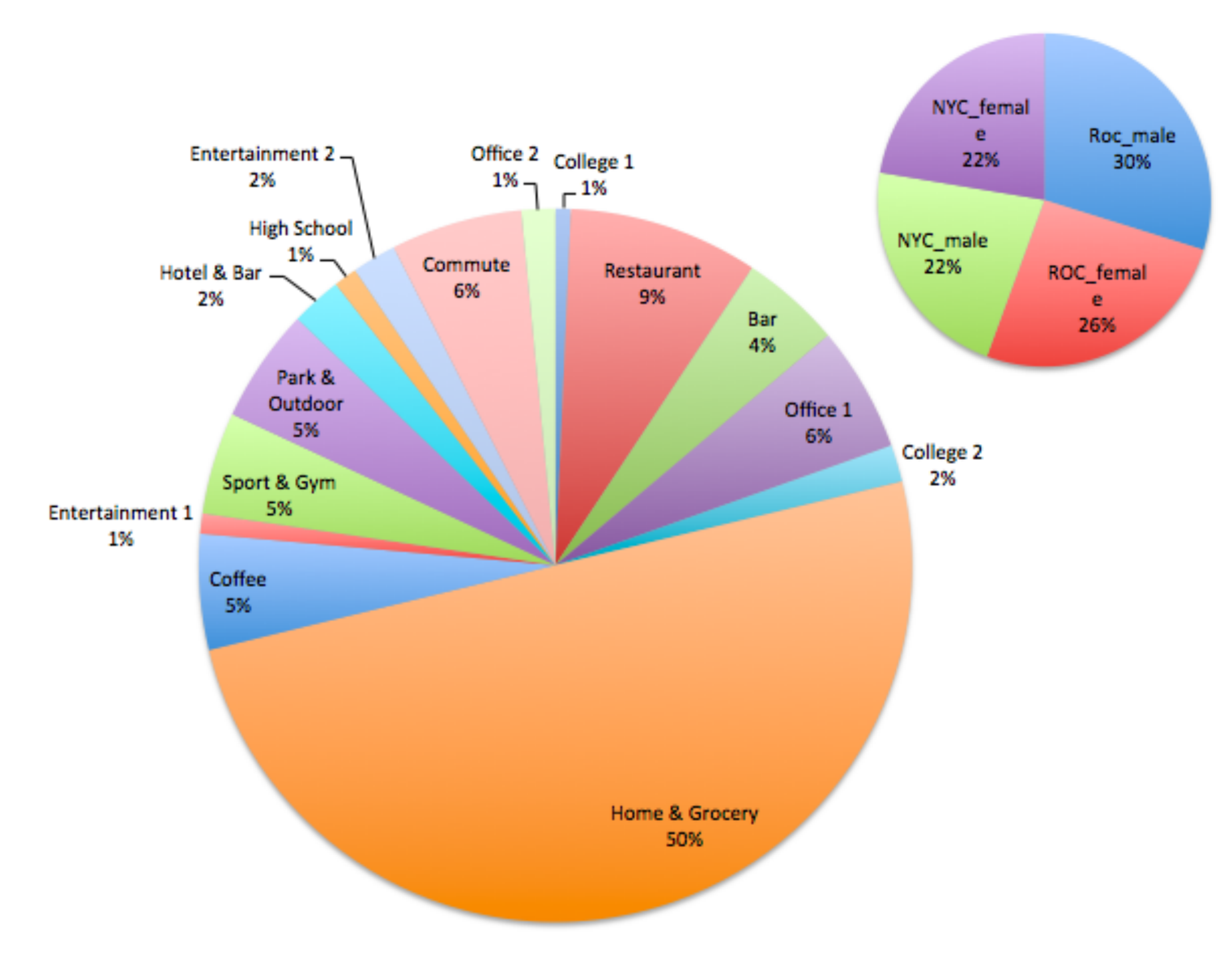}
  \caption{Components and corresponding percentage of lifestyle 1 and the people in this lifestyle.}~\label{fig:lifestyle1}
\end{figure}

\begin{figure}[!htbp]
\centering
  \includegraphics[width=1.0\columnwidth]{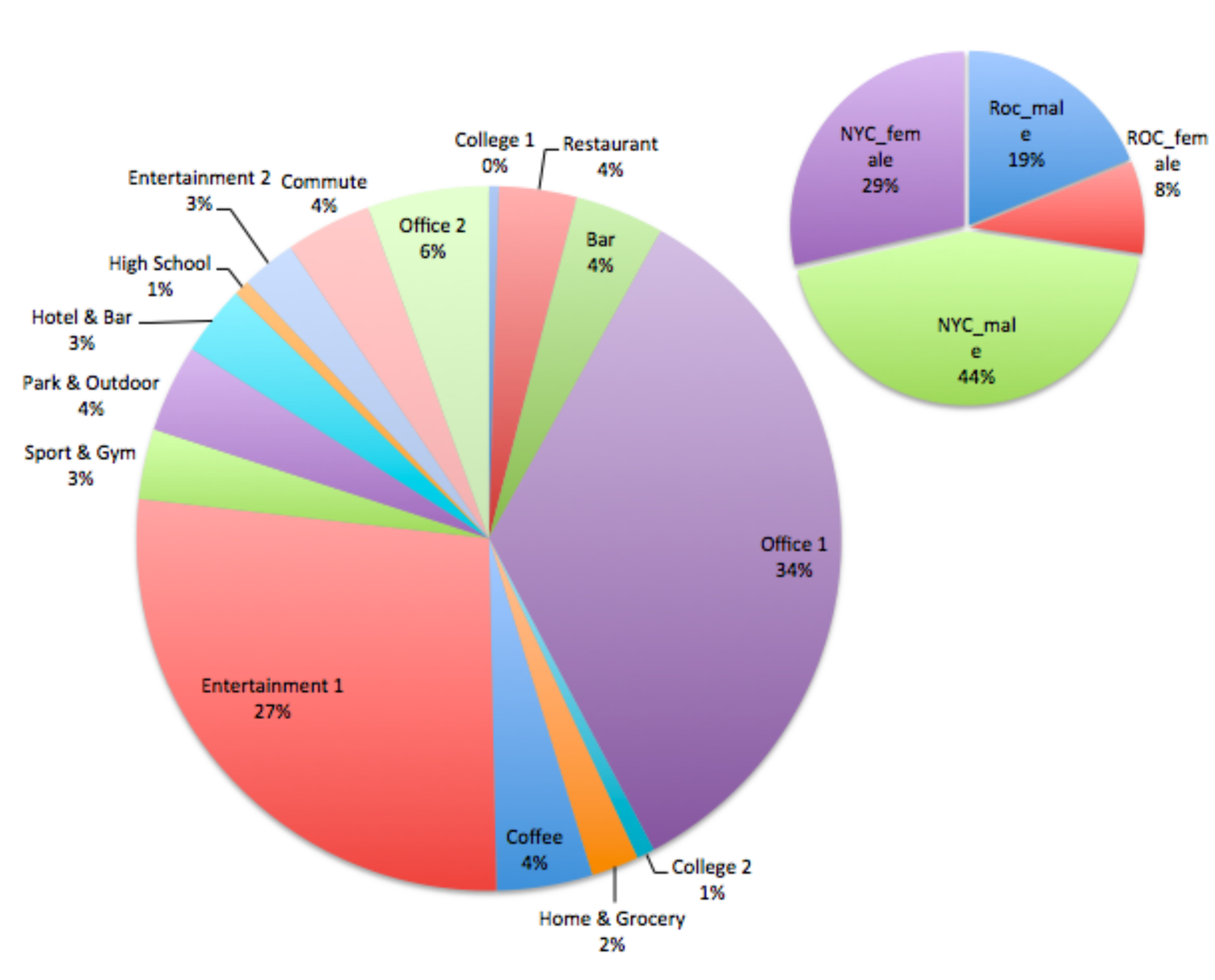}
  \caption{Components and corresponding percentage of lifestyle 2 and the people in this lifestyle.}~\label{fig:lifestyle2}
\end{figure}

We plot the components of two of groups of people and the percentages of males and females from both cities. Note that all ratios are normalized by the number of users in each city. For the people who are in the first group (Fig.~\ref{fig:lifestyle1}), home is the absolute center of their life. Additionally, this group is comprised more of people from small cities (56\%) than of those from large cities (44\%). People in the second group (Fig.~\ref{fig:lifestyle2}) tend to visit office and entertainment venues more often. People in large cities (73\%) prefer this lifestyle more than in small cities (27\%). These results suggest that for people in small cities, home is a prominent location in life; while in large cities, people tend to spend more time at office.

\section{Composite Aspects of Lifestyles}

Some lifestyle patterns may be seen a combination of individuals' daily, weekly, and spatial habits. In this section, we consider the analysis of tensors $T_1$ and $T_2$. Each tensor $T_i\in\mathbb{R}^{N \times M \times P}$ may be factorized into any number of components $k=[2, min\{N,M,P\}]$ . Recall that for both $T_1$ and $T_2$, $N$ indexes check-in counts by user id and $P=100$ indexes by category. For $T_1$, $M=24$ for indexing by time of day, and for $T_2$, $M=7$ for days in a week.

There is a significant trade off that exists with choosing tuning parameter $k$: with a smaller $k$, fewer lifestyle patterns may be identified, and some components may be mixtures of multiple theoretically distinct lifestyle patterns. With higher $k$, we run into issues of redundancy, where multiple highly similar lifestyle patterns are extracted, and low interpretability, where some extracted patterns include very disparate behaviors. For both $T_1$ and $T_2$, we tested a wide range of possible $k$ values. 

Individuals with fewer check-ins than some threshold $h$ were pruned to remove outlier noise. In our experiments, we found very little difference between many components when $h=5$ from when $h=30$. However, with $h=5$, some patterns emerge more apparently, and components are more distinct overall.

\begin{figure}[ht!]
    \includegraphics[width=1.0\columnwidth]{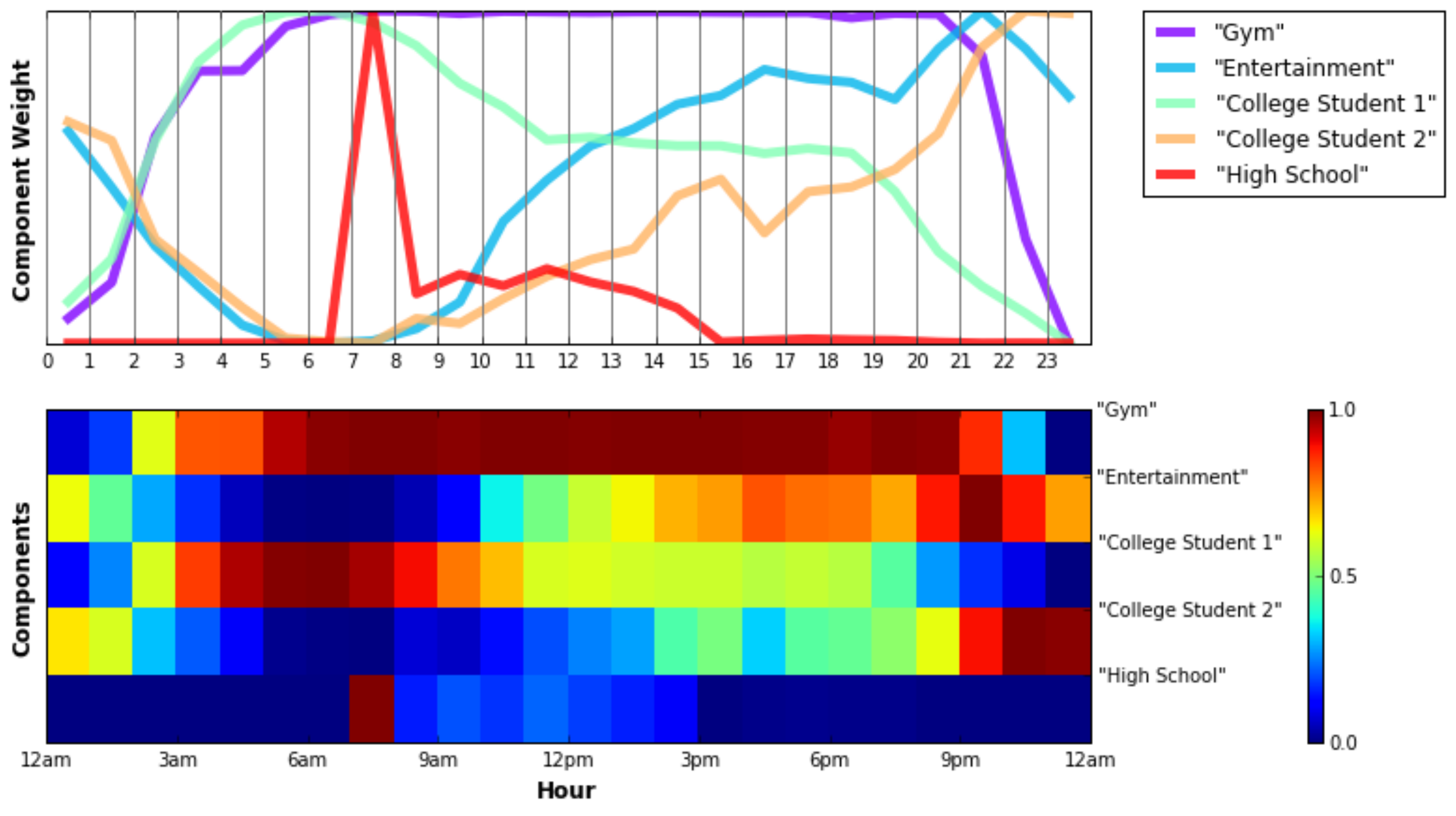}
    \caption{Component weights $L_M$ by hour for tensor $T_1$; 5 out of $k=12$ total components are shown. Component labels are added a posteriori, and weights shown are normalized by min-max normalization. Each curve in the upper part represents the trend of a POI category along 24 hours of a day.}
    \label{fig:T1-plot}
\end{figure}

\begin{figure}[ht!]
    \includegraphics[width=1.0\columnwidth]{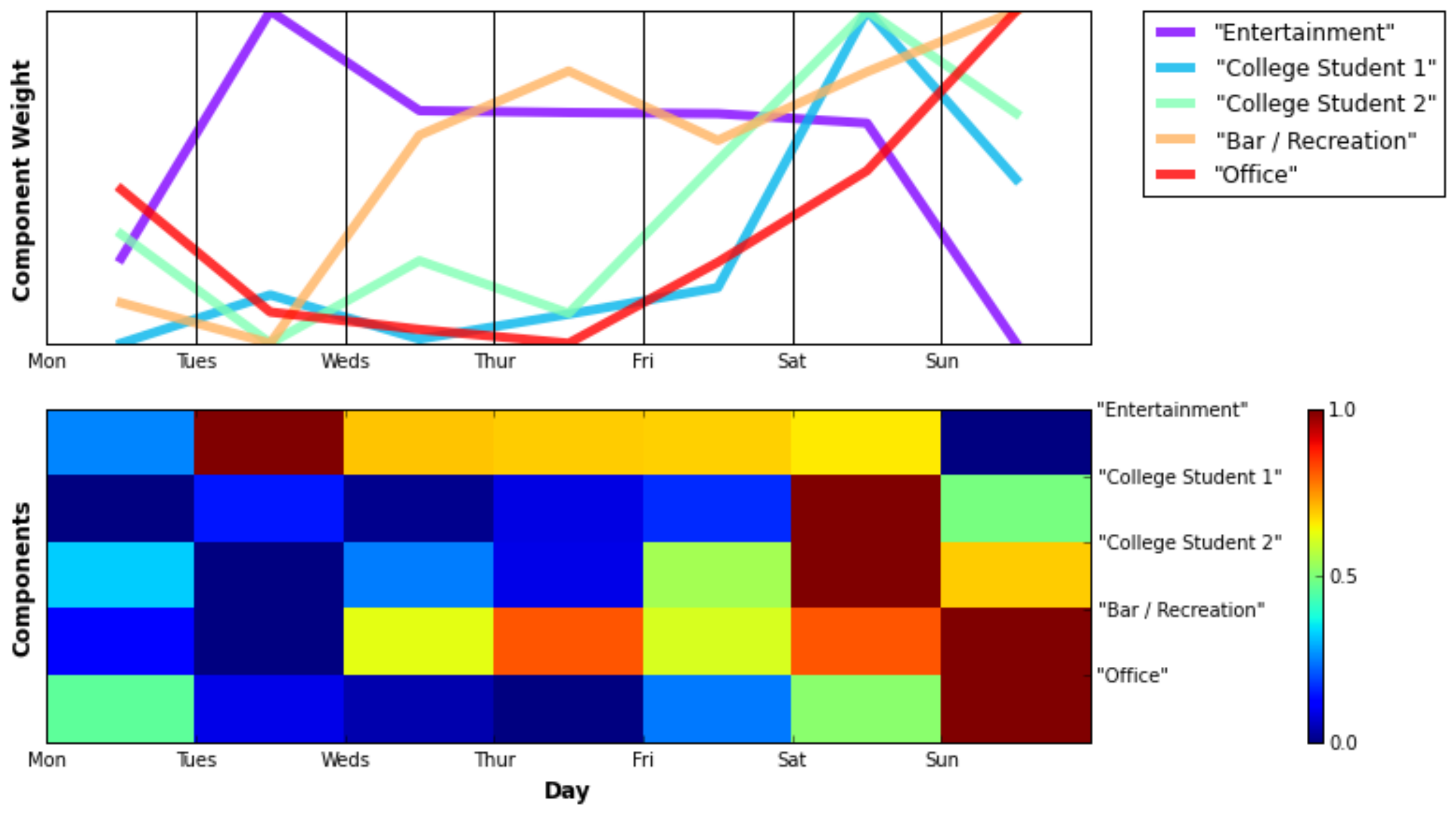}
    \caption{Component weights $L_M$ by day for tensor $T_2$ with $k=5$. Component labels are added a posteriori, and weights shown are normalized by min-max normalization. Each curve in the upper part represents the trend of a POI category along 7 days of a week.}
    \label{fig:T2-plot}
\end{figure}

\subsection{Time-of-Day \& Location}

$T_1$ offers the most interpretable results of third-order tensor decomposition, most clearly when $k=12$, shown in Fig.~\ref{fig:T1-plot}. Across a wide range of values for $k$, we see a few distinct lifestyle patterns emerge. One component assigns highest weight to the Arts \& Entertainment category and significantly lower for all others, with time-of-day beginning around 10am, peaking at 9pm, and tailing off in the hours following midnight. This matches the intuitive assumption that individuals usually visit these sort of venues later in the day, primarily around evening and night times. The next component assigns highest weight to the High School category, significantly lower for all others, with time-of-day peaking significantly at 7am, and some additional weight for the hours of 8am - 3pm. This range directly corresponds to the standard school day for high school students in the U.S. Although NYC check-ins are only collected for the month of June, the school year for New York public schools continues through June 26th.

Two ``college student'' lifestyle patterns consistently emerge. The first assigns the most weight to College Residence Hall, with time-of day gradually peaking around 6am, and decreasing significantly from 8pm through 2am. The second has highest weight on College Rec. Center, with time-of-day increasing gradually from 9am to 9pm, peaking from 10pm to 2am, and tailing off quickly thereafter. Both these lifestyles assign some weight to other college-related POIs, for example College Lab, Co-working Space, and College Cafeteria. The former of these seems to model the pattern ``early bird'' college students, where the latter models ``night owls''.

Finally, we also see a Gym lifestyle emerge, where Gym is assigned a very high weight, and all other categories are assigned low weight. This lifestyle pattern also gives high weight to most hours of the day, with a significant dip from the hours of 10pm through 5am. This also makes intuitive sense, since it is unlikely that many people go to the gym during these hours.

\subsection{Day-of-Week \& Location}

We find noteworthy patterns when decomposing $T_2$ into 5 components, shown in Fig.~\ref{fig:T2-plot}. We see one pattern with highest weight assigned to category Bar, high weights to Cafe, American Restaurant, Private Home, and Music Venue, and moderate weights to a number of similar categories such as Rock Club; this pattern assigns very little weight to Monday and Tuesday, and the highest weight on Sunday. Common sense tells us that people work harder the first few weekdays, and usually visit recreational venues such as these more later in the week. It is not surprising that the highest weight is assigned on a weekend night, since this includes check-ins both the night of that day, and activities past midnight from the night before. 

We also see a component with very high weight assigned to Arts \& Entertainment, low weight assigned to other categories; very low weight is assigned to Monday, no weight to Sunday, and relatively uniform weights across other days of the week. Many entertainment venues are closed on Sunday, and fitting with the common notion that people recreate less on Mondays. Two distinct ``college student'' patterns emerge with as few as 4 components, both with considerably higher weights assigned to weekends than weekdays. It is plausible that college students, especially in NYC, go off-campus to engage in alternate lifestyle behaviors during the weekdays when they're not in class, and stay on campus to study during the weekends.

The fifth and final component we see assigns highest weight to the Office category, and weights an order below to a few categories: American Restaurant, Pub, and Deli. This pattern has a less clear interpretation; one might speculate that individuals who go to the office on weekends might develop a habit of checking into social media outlets during these irregular visits, but not during their daily grind during the weekdays.

\section{Conclusion and Future Work}
In this paper, we extensively study the differences between lifestyles of a big city (NYC) and of a smaller city (ROC) using social media data. We extract work-rest habits and lifestyles from user activities. Instead of assigning people to qualitatively defined work-rest classes, we apply NMF techniques to discover latent patterns of human diurnal preference. The extracted latent patterns correspond well to the intuitively defined classes. Also using NMF, we find hidden features of human movement preference. We then group residents of two cities into lifestyle clusters based on the weights of hidden features and analyze the difference between the two distinctive cities. Moreover, tensor decomposition techniques are applied to find composite life patterns in our work. Clear and quantifiable differences are found in the lifestyles of large and small cities. 

Lifestyle is a broad, imprecise concept that covers a multitude of aspects of human behavior, and social media is a flawed representation of individuals' daily behaviors. These challenges present a number of exciting avenues for future work: investigating what sorts of lifestyles can be categorized culturally, or by other sociological factors such as job and age group; establishing what behaviors can be more characteristic of specific lifestyles; and exploring the relationship of individuals' social media activities to their daily behaviors. Regarding the last point, it may be that certain individuals' social media postings are strongly indicative of their behaviors, whereas other individuals' postings may be a biased sample of their activities.

We would like to introduce more dimensions in our future work, such as other demographic dimensions including income, age and race, as well as adding more cities to the investigation. Presently, we are collecting data from the San Francisco Bay Area so that we may compare lifestyle behaviors between inhabitants of east and west coast cities, and relate our findings to previous research ~\cite{fincher1998cities}. We also plan to combine multiple social media data sources to gain a comprehensive understanding of human behaviors in the big data era using large-scale social media data. 

\section{Acknowledgment}
We would like to thank the support from the New York State through the Goergen Institute for Data Science, as well as Xerox Foundation.

\balance{}

\bibliographystyle{IEEEtran}

\bibliography{reference}

\begin{IEEEbiography}[{\includegraphics[width=1in,height=1.25in,clip,keepaspectratio]{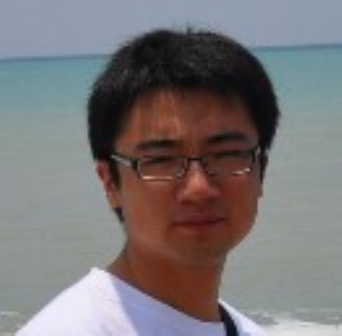}}]{Tianran Hu}
is a second year PhD student in the Computer Science Department of University of Rochester, advised by Prof. Jiebo Luo.  He obtained his M.S. degree in 2012 from Hong Kong University of Science and Technology and B.Eng. in 2010 from Sichuan University in China. His research interests are in social media and data mining. 
\end{IEEEbiography}

\begin{IEEEbiography}[{\includegraphics[width=1in,height=1.25in,clip,keepaspectratio]{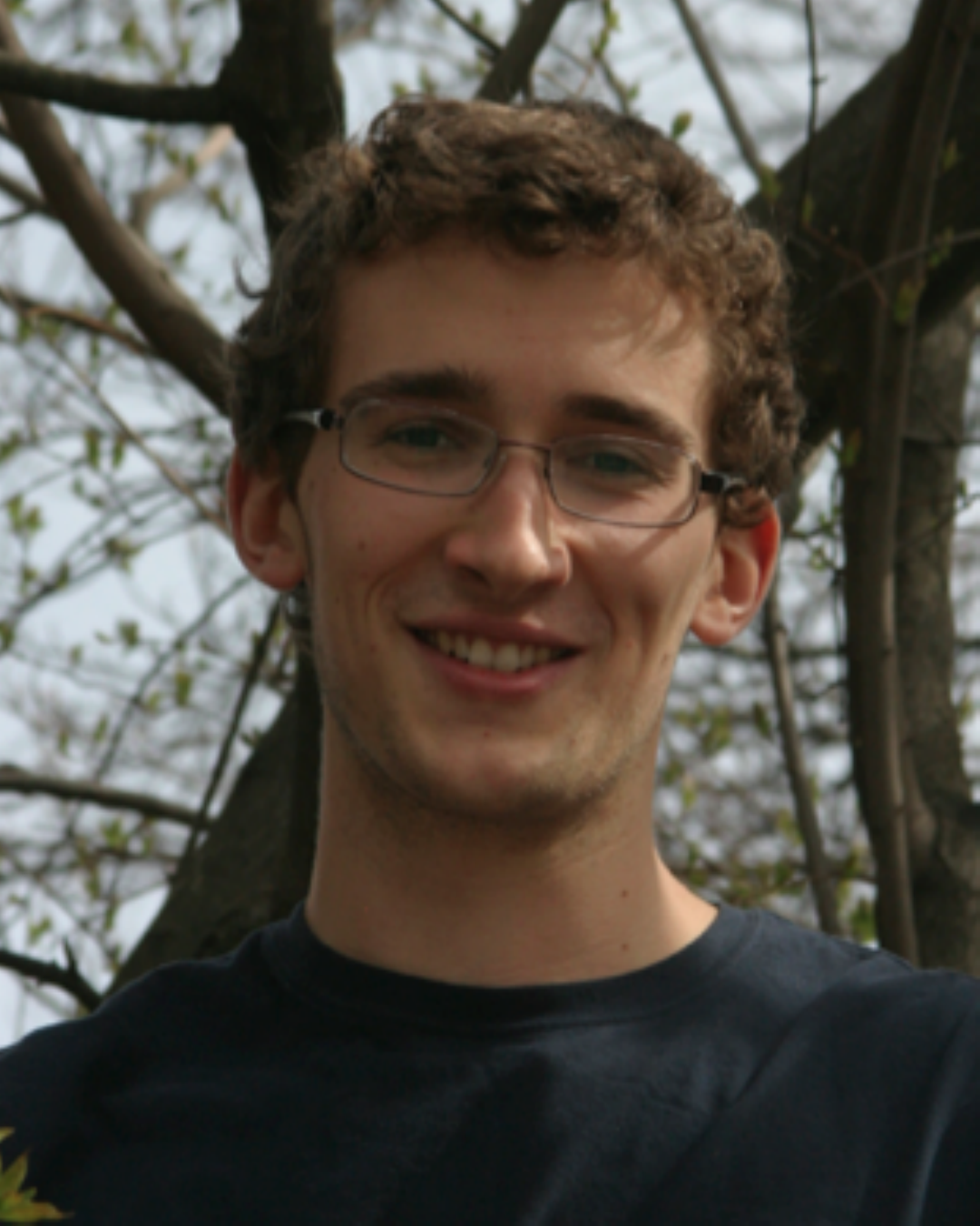}}]{Eric Bigelow}
is a first year M.S. student in the Computer Science Department at the University of Rochester. Previously, he obtained a B.S. in Brain \& Cognitive Sciences and a B.A. in Computer Science from the University of Rochester in 2014. His research interests span applications and theoretical foundations of machine learning.

\end{IEEEbiography}

\begin{IEEEbiography}[{\includegraphics[width=1in,height=1.25in,clip,keepaspectratio]{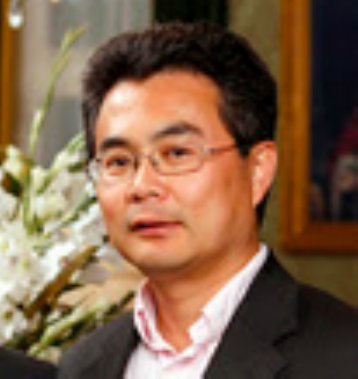}}]{Jiebo Luo}
joined the University of Rochester in Fall 2011 after over fifteen years at Kodak Research Laboratories, where he was a Senior Principal Scientist leading research and advanced development. He has been involved in numerous technical conferences, including serving as the program co-chair of ACM Multimedia 2010 and IEEE CVPR 2012. He is the Editor-in-Chief of the Journal of Multimedia, and has served on the editorial boards of the IEEE Transactions on Pattern Analysis and Machine Intelligence, IEEE Transactions on Multimedia, IEEE Transactions on Circuits and Systems for Video Technology, Pattern Recognition, Machine Vision and Applications, and Journal of Electronic Imaging.  Dr. Luo is a Fellow of the SPIE, IEEE, and IAPR.
\end{IEEEbiography}

\begin{IEEEbiography}[{\includegraphics[width=1in,height=1.25in,clip,keepaspectratio]{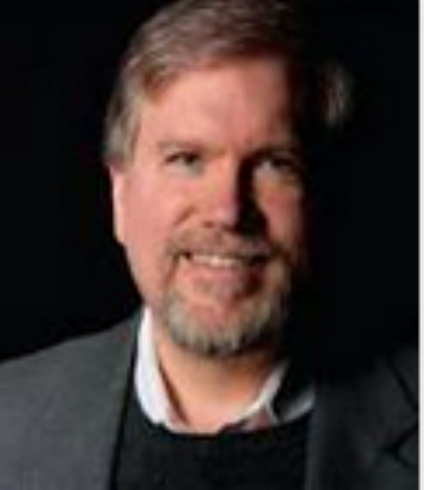}}]{Henry Kautz}
is the Robin \& Tim Wentworth Director of the Goergen Institute for Data Science and Professor in the Department of Computer Science at the University of Rochester. He has served as department head at AT\&T Bell Labs in Murray Hill, NJ, and as a full professor at the University of Washington, Seattle. In 2010 he was elected President of the Association for Advancement of Artificial Intelligence (AAAI). His research in artificial intelligence, pervasive computing, and healthcare applications has led him to be honored as a Fellow of the American Association for the Advancement of Science (AAAS), Fellow of the Association for Computing Machinery (ACM), and Fellow of the AAAI. He holds the Computers \& Thought Award from the International Joint Conference on Artificial Intelligence. In 2013, he received the Ubicomp 10-Year Impact award for his paper "Inferring High-level Behavior from Low-level Sensors".
\end{IEEEbiography}

\end{document}